\documentclass[pra,letterpaper,twocolumn]{revtex4-1}
\usepackage[utf8x]{inputenc}
\usepackage{amsmath}
\usepackage{xcolor}%
\usepackage{graphicx}% Include figure files
\usepackage{dcolumn}% Align table columns on decimal point
\usepackage{bm}% bold math
\usepackage{subfigure}
\usepackage{textcomp}

\newcommand{\ket}[1]{\left| #1 \right\rangle}
\newcommand{\bra}[1]{\left\langle #1 \right|}

\newcommand{\abs}[1]{\left| #1 \right|}

\newcommand{\expn}[1]{{\rm e}^{#1}}
\newcommand{\bv}[1]{\mathbf{#1}}
\newcommand{\dg}{{^{\dagger}}}

\newcommand{\ie}{\textit{i.e.,}~}
\newcommand{\eg}{\textit{e.g.,}~}

\newcommand{\etal}{{\emph{et al.}~}}

\newcommand{\bvN}{\bv{N}}
\newcommand{\bvM}{\bv{M}}

\begin{document}
	\title{\textbf{General Modeling Framework for Quantum Photodetectors}}
	\author{Steve M. Young}
	\author{Mohan Sarovar}
	\author{Fran\c{c}ois L\'{e}onard}
	\affiliation{Sandia National Laboratories, Livermore, CA, 94551, USA}

	\begin{abstract}
Photodetection plays a key role in basic science and technology, with exquisite performance having been achieved down to the single photon level. Further improvements in photodetectors would open new possibilities across a broad range of scientific disciplines, and enable new types of applications. However, it is still unclear what is possible in terms of ultimate performance, and what properties are needed for a photodetector to achieve such performance. Here, we present a general modeling framework for photodetectors whereby the photon field, the absorption process, and the amplification process are all treated as one coupled quantum system. The formalism naturally handles field states with single or multiple photons as well as a variety of detector configurations, and includes a mathematical definition of ideal photodetector performance.  The framework reveals how specific photodetector architectures introduce limitations and tradeoffs for various performance metrics, providing guidance for optimization and design. 
	\end{abstract}
	\maketitle

\section{Introduction}

Photodetectors are used extensively in a broad range of scientific experiments and for numerous technology applications. Pushing the limits of photodetection is important to allow new phenomena to be explored and to improve the performance in applications. In the realm of single-photon detection, records are constantly being reported for detector performance \cite{Bienfang:2004ij,Woodson:2016cx,Pernice:2012bc,Marsili:2012ib,Marsili:2013th,Eisaman:2011cc, Hadfield:2009}, and these detectors
are being used to gain increasing understanding of the the fundamental properties of light \cite{Cassemiro:2010, Genovese:2017, Perina:2003, Achilles:03, Wamsley:2016}. Still, photodetectors are complex, and determining the fundamental limits of their performance and how to design their internal structure is not straightforward because of several simultaneous performance requirements. In the case of single photon detectors, these performance metrics include efficiency, jitter, dark count rate, number resolution, and bandwidth, but also include aspects such as operating temperature, size, and power requirements. Determining the best possible photodetector that can optimize all of these considerations is challenging because many of these are interrelated.

Since the formulation of quantum mechanics there have been several theoretical models of photodetection. The pioneering work of Glauber \cite{glauber_1963}, Mandel \etal \cite{mandel_1964}, and Kelly and Kleiner \cite{kelley_1964} established the relationship between the counting statistics of point-like detectors and states of the electromagnetic field, and this theory has provided the foundation for much of the proceeding work. Refinements of the theory by Scully and Lamb \cite{scully_quantum_1969}, Srinivas and Davies \cite{srinivas_photon_1981} and Ueda \etal \cite{ueda_quantum_1990} accounted for the backaction of the detection process on the field, which is important to capture the statistics of continuous photocurrents in the limit of weak fields. Further important refinements of photodetection theory include the relaxation of approximations in the field-matter interaction, \eg the rotating-wave approximation \cite{drummond_unifying_1987, fleischhauer_quantum-theory_1998}, and the incorporation of variations in detector architecture, \eg multiplexed arrays \cite{sperling_true_2012}.

While the theoretical models and methods resulting from this large body of literature are useful for understanding photodetection phenomena in many contexts, they generally do not provide a framework to design photodetectors from the ground up. Such an endeavor might have been experimentally unfeasible in the past, but with recent progress in nanoscale fabrication and engineering one can now ask the question of how to design an optimal photodetector starting from the atomic scale. To answer this question one needs a theory that models the dynamics of the electromagnetic field and some general model of the detector's internal degrees of freedom, with as few assumptions as possible. Such a design-oriented approach is essential to establish the ultimate limits of photodetectors, and to identify from general principles new optimal designs, or perhaps even radically new photodetector designs. For example, recent progress in developing such an approach to modeling photodetectors has shown that in principle there exists no trade-offs between some detector metrics \cite{vanEnck:2017, Young:2018}, suggesting that improved photodetectors are possible.

In this manuscript we build on theories for light-matter interaction with weak fields, open quantum systems, and quantum measurement, to develop a holistic approach to modeling photodetectors that allows one to directly relate general criteria for performance to internal photodetector structure, and moreover, optimize this internal structure to meet performance metrics. Our formalism allows one to evaluate the state of the matter system during and after interaction with the field as
\begin{flalign}
&\hat{\rho}_{\rm MATTER}(t) = {\rm Tr}_{\rm LIGHT}\left[\mathcal{P}(t,t_0)\hat{\rho}_{\rm TOT}(t_0)\right]\label{eq:basic}
\end{flalign}
where $\mathcal{P}$ is an operator determined by the internal structure of the system and its coupling to both the incident field and amplification processes, and $\hat{\rho}_{\rm TOT}(t_0)=\hat{\rho}_{\rm LIGHT}(t_0)\otimes\hat{\rho}_{\rm MATTER}(t_0)$ represents the initial density operator for the combined matter and field quantum state. ${\rm Tr}_{\rm LIGHT}$ represents a partial trace over the field degrees of freedom. We show how to explicitly calculate $\mathcal{P}$ in a wide variety of cases, and additionally, we show that in most cases of interest, this allows us to represent measurement outcomes $\Pi(t)$ as 
\begin{flalign}
&\Pi(t) = {\rm Tr}_{\rm LIGHT}\left[\mathcal{K}(t,t_0)\hat{\rho}_{\rm TOT}(t_0)\right]\label{eq:basic2}
\end{flalign}
which can be used to determine average performance.
This abstraction enables us to both intuitively understand detection as propagation of an input pulse to an outgoing signal and analyze the effects of detector internal structure on average photodetector performance. In particular it naturally furnishes a definition for ideal detection that places conditions on $\mathcal{P}$ (or $\mathcal{K}$) -- and, consequently, the detector architecture -- that must be satisfied to achieve it, ultimately allowing us to identify new photodetector designs with superior predicted performance. 

\begin{figure}
	\includegraphics[width=\columnwidth]{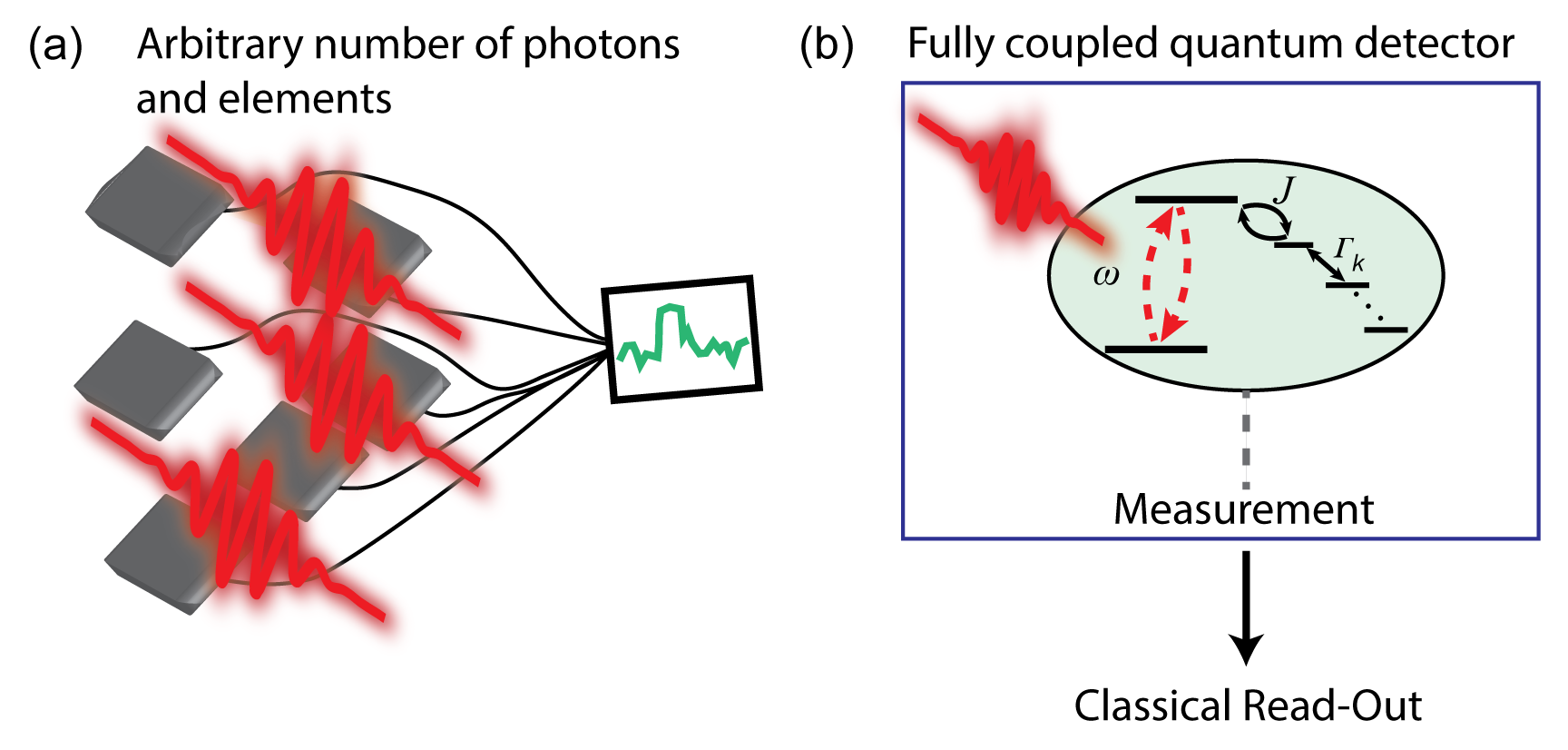}
	\caption{Illustration of the general photodetection situation being considered. (a) An arbitrary number of photons impinges on an arbitrary number and configuration of elements. An element is an object that
	creates a signal upon photon absorption. It can consist of a few atoms, or 
macroscopic collections of atoms.(b) A fully coupled quantum detector considers the photon field, the optical absorption, and the measurement as part of one quantum system. The result of the measurement is fed to a classical read-out. \label{fig:system}}
\end{figure}

\section{Modeling Approach}
\label{sec:ma}

We seek to develop an approach that can describe a general photodetection system as illustrated in Fig.~\ref{fig:system}. By a general photodetection system we mean one that is composed of a number
of elements, where the function of each element is to generate a signal upon photon absorption. An element
could be composed of a few atoms, or could be microscopic in size. For example, a single molecule on a
few-atom quantum transport channel could be an element; a macroscopic semiconductor composed of interacting atoms could also be an element. In both cases one signal is generated. The general photodetection system is 
composed of many such elements, each interacting with the field, and each generating its own signal. The only
limitation we impose on the photodetection system is that it is smaller than the photon wavelength.
 
In Ref.~\cite{Baragiola:2012cs} it was shown that the Ito Langevin equation for the interaction of a point-like (\ie within a photon wavelength) matter system with a few-photon pulse can be solved by rewriting it as a set of master equations constructed according to the initial incoming field in the Fock basis. In this manuscript we consider a single-mode
wavepacket, which is relevant to detectors in single-mode waveguides, but the approach can be generalized to more
complex fields. For a single-field-mode wavepacket with frequency $\omega_E$ and light-field temporal profile at the detector of $E(t)$, the overall density matrix for the field is
\begin{flalign*}
\hat{\rho}_{\rm LIGHT}(t) = \sum_{N,M} c_{N,M}(t)\ket{N}\bra{M},
\end{flalign*} 
where $\ket{N}$ is a Fock state of the mode with occupation $N$, see Appendix \ref{sec:lightfield}. We note that in the case of specific numerals in place of $N$ and $M$ separating commas will be omitted.
For a field initially in a state defined by a set of $c_{N,M}(t_0)$ the reduced density matrix characterizing the matter degrees of freedom at any time $t$ is 
\begin{flalign*}
&\hat{\rho}_{\rm MATTER}(t) = \sum_{N,M} c_{N,M}(t_0)\hat{\varrho}^{N,M}(t)
\end{flalign*}
where $\hat{\varrho}^{N,M}(t)$ are a set of auxiliary density matrices obeying the equations 
\begin{flalign}
\dot{\hat{\varrho}}^{N,M}(t)&=\mathcal{V}_{\rm SYS}+\mathcal{V}_{\rm F-M}+\mathcal{V}_{\rm AMP}
\nonumber\\
\label{eq:hme}
\end{flalign}
with
\begin{flalign}
\mathcal{V}_{\rm SYS}&=-i[\hat{H},\hat{\varrho}^{N,M}(t)]+\sum^{\rm BATHS}_i\mathcal{D}[\hat{Y}_i]\hat{\varrho}^{N,M}(t),\nonumber\\
\mathcal{V}_{\rm L-M}&=\sqrt{N}E(t)\expn{-i\omega t}[\hat{S}\hat{\varrho}^{N-1,M},\hat{L}^\dagger_{i}]\nonumber\\
&\hspace{0cm}+\sqrt{M}E^*(t)\expn{i\omega t}[\hat{L}_{i},\hat{\varrho}^{N,M-1}\hat{S}^\dagger]+\mathcal{D}[\hat{L}]\hat{\varrho}^{N,M}\nonumber\\
&+\sqrt{MN}\abs{E(t)}^2\left(\hat{S}\hat{\varrho}^{N-1,M-1}\hat{S}^\dagger-\hat{\varrho}^{N-1,M-1}\right),\nonumber\\
\mathcal{V}_{\rm AMPS}&=\sum^{\rm AMPS}_i\mathcal{D}[(2k_i)^{1/2}\hat{X}_i]\hat{\varrho}^{N,M}(t),
\label{eq:full_me}
\end{flalign}
with ``BATHS'' and ``AMPS'' referring to sums over the number of baths and amplification channels, respectively.  It is straightforward, if cumbersome, to extend this formalism to multiple field modes, including additional spontaneous emission channels.  

Eq.~\eqref{eq:hme} comprises three parts: the internal dynamics of the system, the light-matter interaction due to the incoming field excitation, and the amplification of internal states of the detector, modeled as a weak measurement of some internal states. $\mathcal{V}_{\rm SYS}$ describes the internal evolution according to $\hat{H}$, the non-interacting Hamiltonian, and the influence of external baths $\hat{Y}_i$, described using the Lindblad superoperator 
\begin{flalign*}
\mathcal{D}[\hat{O}]\hat{\rho}=\hat{O}\hat{\rho} \hat{O}^\dagger-\frac{1}{2}\hat{O}^\dagger \hat{O}\hat{\rho} -\frac{1}{2}\hat{\rho} \hat{O}^\dagger \hat{O}.
\end{flalign*} 

$\mathcal{V}_{\rm L-M}$ describes the light-matter interaction, including spontaneous emission, and is mediated by the dipole coupling $\hat{L}$ and the quadratic coupling $\hat{S}$. It is important to note that these terms dictate that the evolution of a given auxiliary density matrix relies on the evolution of auxiliary density matrices interacting with fields with reduced initial excitations.   
\begin{figure}
	\includegraphics[width=\columnwidth]{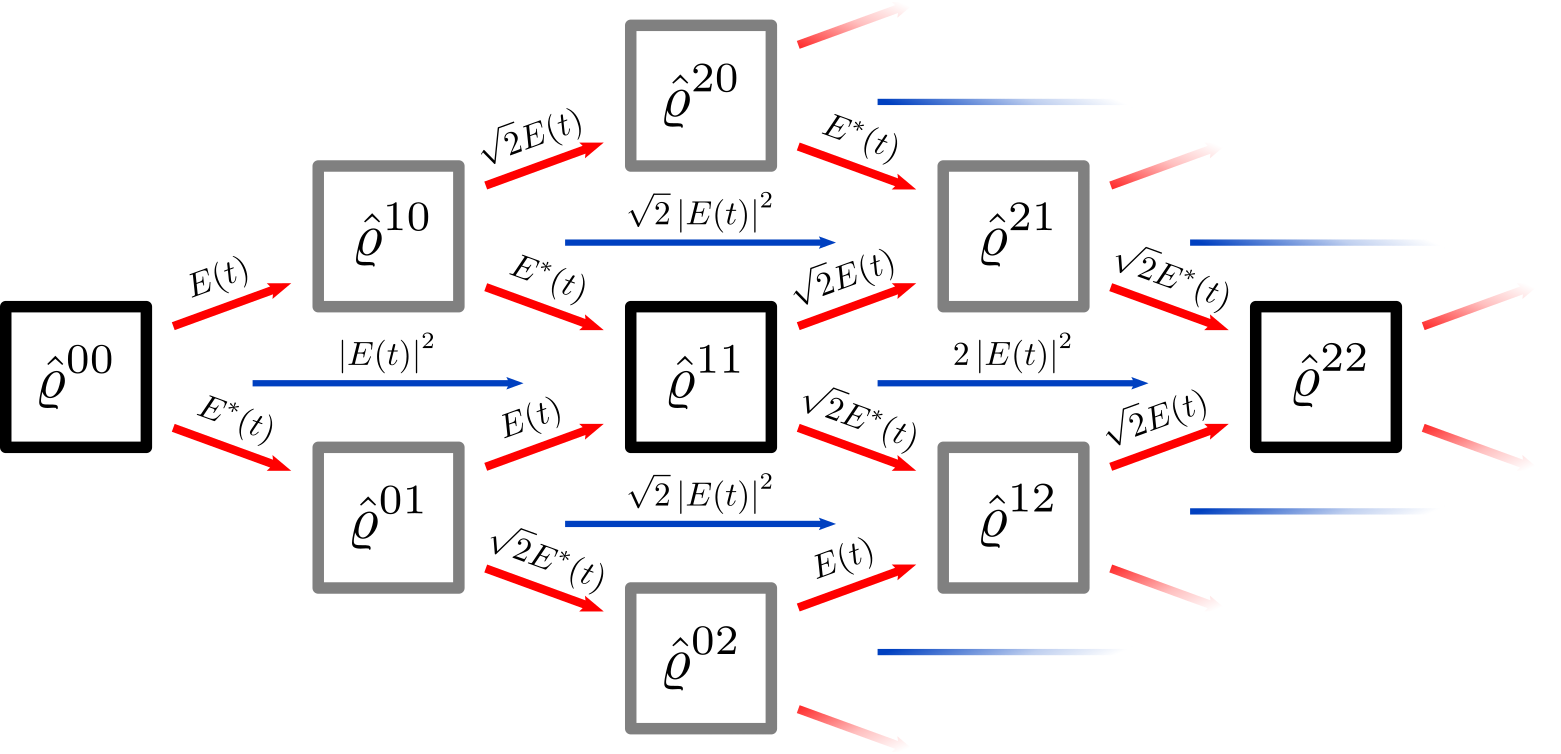}
	\caption{The relationship between matrices $\hat{\varrho}^{N,M}(t)$ evolved by the system of equations in \eqref{eq:hme}.  Each matrix depends on the evolution of matrices with lower superscript index; the real system density matrix is given by the matrix with superscript indices corresponding to the initial state of the incoming field.  \label{fig:block_rho}}
\end{figure}
We emphasize that these field indices do not directly correspond to the state of the field at a time $t$; they correspond to the state of an incoming field before interaction at time $t_0$, and the density matrices that contain the correct information about the system are the ones that correspond to the actual incoming field state given by $c_{N,M}(t_0)$.  So, for an initial field with Fock state in a single mode containing two photons, the density matrix that contains the true dynamics of the system is $\hat{\varrho}^{22}$.  The others are auxiliary density matrices required to propagate $\hat{\varrho}^{22}$ and are in some sense fictitious.  This is shown schematically in Fig.~\ref{fig:block_rho}.  Each diagonal density matrix ($N=M$) is initialized to the state of the system ($\hat{\varrho}^{N,N}(t_0)=\hat{\rho}_{\rm MATTER}(t_0)$), and the off-diagonal ones ($N \ne M$) are zero.   

The amplification process $\mathcal{V}_{\rm AMPS}$ is mediated by operators $\hat{X}_i=\sum_{j}\chi_{ij}\hat{x}_{ij}$, which we constrain to be Hermitian and therefore expressible as sums of projectors onto internal states $\hat{x}_{ij}=\ket{v_j}\bra{v_j}$. These operators also influence the dynamics through the Lindbladian.
These operators can be interpreted as a weak measurement process with amplification strengths $\chi_{ij}$ and rate $k_i$\cite{Jac.Ste-2006}.  However, Eq.~\eqref{eq:hme} as written only describes the averaged dynamics associated with amplification. It may be unraveled\cite{Jac.Ste-2006} into individual measured trajectories by conditioning the dynamics on the prior results of the measurement imposed by the amplification.  This is accomplished by adding the term 
\begin{flalign}
 \sum^{\rm AMPS}_i\frac{dW_{i,t}}{dt}(2k_i)^{1/2}\left(\hat{X}_i\hat{\varrho}^{N,M}+\hat{\varrho}^{N,M}\hat{X}_i^\dagger-2\langle \hat{X}_i\rangle \hat{\varrho}^{N,M}\right) \label{eq:amp}
\end{flalign}
to Eq.~\eqref{eq:hme}, where $W_i$ are Wiener processes for each amplification channel that correspond to particular measurement records, whose increments are explicitly,
\begin{flalign}
dI_i(t)= \langle \hat{X}_i\rangle(t)dt+\frac{1}{(8k)^{1/2}}dW_{i,t}.
\end{flalign}
We assume the Wiener processes across measurement channels are uncorrelated, and for later use define the integrated measurement records
\begin{flalign}
I_i(t)=\frac{1}{t_m}\int^t_{t-t_m}dt'\left(\langle \hat{X}_i\rangle(t')+\frac{1}{(8k)^{1/2}}\frac{dW_{i,t'}}{dt'}\right)\label{eq:record}
\end{flalign}
where $t_m$ defines an integration time that sets the temporal resolution of the detector. As an example, $I_i(t)$ could be the time-dependent current measured with an external electronic read-out. We note the appearance of the quantity $\frac{1}{(8k)^{1/2}}\frac{dW_{i,t}}{dt}$, which is noise arising from the continuous measurement process.

The machinery introduced thus far allows for determination of the average and unraveled dynamics of the system, and in principle, computation of measurement outcomes.  However, it cannot be written in the form of the direct relations that we desire as expressed by Eqs. \eqref{eq:basic} and \eqref{eq:basic2}.  We will now restructure and extend this machinery so that we may write the system dynamics and performance in this form, considerably enhancing its power and utility.  

The system of equations in Eq.~\ref{eq:hme} is linear (and inhomogeneous), since it may be solved in sequence and the time-dependent field modifies only the input arising from previously solved systems of equations. However, solving these differential equations in this form is inconvenient; they are expressed in terms of rank 4 linear operators, instead of the typical matrix operators. It is therefore preferable to recast these equations by rewriting, \eg each $\hat{\rho}$ as a vector $\bar{\rho}$, the internal and amplification superoperators as matrices, and the coupling to other density matrices as an input vector.  A common scheme for this conversion is to take $\bar{\varrho}_{i\cdot n+j}=\hat{\varrho}_{j,i}$; in this case superoperators may be converted using the relation $\bar{(\hat{O}\hat{\rho}\hat{Q})}=(\hat{Q}^{\rm T}\otimes\hat{O})\bar{\rho}$, where $\otimes$ is the Kronecker product. However, this transformation is not necessarily optimal, and we will employ others as is convenient.  
For evaluation of the output records from Eq.~\ref{eq:record} the amplification operators $\hat{X}_i$ become $\bar{X}_i=\sum_j\chi_{ij}\bar{x}_{ij}$, where $\bar{x}_{ij}$ is the vectorized form of $\hat{x}_{ij}$ and $\langle\hat{X}_i\rangle(t)=\bar{X}_i\cdot \bar{\rho}_{MATTER}(t) = \bar{X}_i\dg \bar{\rho}_{MATTER}(t)$. For later convenience we also define $\hat{x}_i = \sum_j \hat{x}_{ij}$ and its vectorized form $\bar{x}_i$, which are projectors onto the monitored internal states.

Thus we have for the average dynamics
\begin{flalign*}
\dot{\bar{\varrho}}^{N,M}(t)&=\bar{\mathcal{A}}\bar{\varrho}^{N,M}(t)+\bar{\beta}^{N,M}(t)
\end{flalign*}
with
\begin{flalign*}
\bar{\beta}^{N,M}(t)&=\abs{E}^2\sqrt{MN}\bar{\mathcal{S}}\bar{\varrho}^{N-1,M-1}(t)\\
&\hspace{2cm}+\expn{-i\omega t}E(t)\sqrt{M}\bar{\mathcal{L}}^+\bar{\varrho}^{N,M-1}(t)\\
&\hspace{2cm}+\expn{i\omega t}E^*(t)\sqrt{N}\bar{\mathcal{L}}^-\bar{\varrho}^{N-1,M}(t).
\end{flalign*}
Here $\bar{\mathcal{A}}$ contains the coefficients of the matrix elements as written above, the matrices $\bar{\mathcal{S}}$ and $\bar{\mathcal{L}}^\pm$ contain the field couplings to density matrices with lower mode indices. The use of $+$ and $-$ superscripts in $\bar{\mathcal{L}}^\pm$ is intended only to distinguish the two objects rather than denote any operation.  Importantly, $\bar{\mathcal{A}}$ contains the information on the internal structure of the photodetector. In the following we will assume that $\bar{\mathcal{A}}$ is time-independent for simplicity. 
Assuming the system is in an initial state $\bar{\rho}(t_0)$ before the wavepacket arrives, the solution is
\begin{flalign}
\bar{\varrho}^{N,M}(t)&=\delta_{N,M}\expn{\bar{\mathcal{A}}(t-t_{0})}\bar{\varrho}^{N,M}(t_0)+\int_{t_0}^td\tau\expn{\bar{\mathcal{A}}(t-\tau)}\bar{\beta}^{N,M}(\tau).\label{eq:sol}
\end{flalign}
Unless otherwise noted, we will assume that $\bar{\varrho}(t_0)$ is an eigenstate of $\bar{\mathcal{A}}$ (e.g., a ground state), 
so that $\expn{\bar{\mathcal{A}}(t-t_{0})}\bar{\varrho}^{N,M}(t_0)=\bar{\varrho}^{N,M}(t_0)$,
and take $t_0=-\infty$ for convenience.
Computing $\bar{\varrho}^{N,M}(t)$ iteratively from $\bar{\varrho}^{00}(t_0)$, we can see that our final expression is a set of nested integrals propagating the initial system state through successive interactions with the field. While this may appear similar to a perturbative expansion in an interaction with a semiclassical field, we emphasize that the physical meaning of the series is distinct.  The computed evolution is exact, having been derived from the complete evolution of the combined, fully quantized light-matter system.  For each interaction the field is implicitly modified to account for changes in the field occupation arising from the interaction. Thus, the expansion terminates due the finite occupation of the field, rather than as an approximation. We note that the only approximation made thus far, beyond the treatment of the detector as point-like, is regarding a timescale separation between the degrees of freedom of the detector that interact with light and the ones that carry the amplified information. This enables us to make the Markov approximation, and describe the amplification as a weak measurement. Crucially, no timescale separation is assumed between the field dynamics and initial detector states, as is typical in traditional photodetection theory \cite{glauber_1963}.

Writing $\bar{\mathcal{G}}(t-\tau)=\expn{\bar{\mathcal{A}}(t-\tau)}$, we note that $\bar{\mathcal{G}}(t-\tau)$ propagates the input from the field-system interaction at $\tau$  to $t$, and thus defines a set of Green's functions that are independent of the field degrees of freedom. Therefore, they characterize the internal modes and amplification of our detector.  In particular, the elements of $\bar{\mathcal{G}}$ will take the form of sums of exponentials with eigenvalues of $\bar{\mathcal{A}}$ as factors of time in the arguments.  Thus the eigenvalues of $\bar{\mathcal{A}}$ play an important role in characterizing the detector performance as we will see below.

In many cases, $\bar{\mathcal{A}}$, and therefore $\bar{\mathcal{G}}(t)$, will be block diagonal, and $\bar{\mathcal{L}}^\pm$ will be block off-diagonal.  This occurs when elements coupled by the field interaction are not coupled by internal processes. As a result, states within blocks will only evolve under the action of $\bar{\mathcal{G}}(t)$ within their blocks, while $\bar{\mathcal{L}}^\pm$ will map one block of states to another.  This can simplify both analysis and solution significantly. 

We note that this Green's function formalism can be related to the conventional POVM formulation of a quantum measurement on the incoming field, see Appendix~\ref{sec:povm}.

\section{Performance}
To obtain detector performance metrics, typically the stochastic master equation consistent with Eqs.~\eqref{eq:sol} (this is explicitly given in Appendix~\ref{sec:eff_estimation}) and the measurement record in Eq. (\ref{eq:record}) must be numerically integrated. These are stochastic trajectories of the system, and thus one has to resort to Monte Carlo averaging over these trajectories to obtain average detector metrics, which can be expensive and cumbersome. However, in many practical parameter regimes it is possible to exploit the properties of Green's functions to obtain estimates of these metrics using only the average evolution equation, Eq.~\eqref{eq:sol}. 

In particular, we desire an expression for the probability $\Pi_{i}(N,t)$ that at time $t$ the $i$th channel has recorded a hit for an incoming field with $N$ photons. 
A hit is recorded when the output $I_i(t)$ exceeds a threshold $I_{HIT,i}$.
In the strong amplification regime, the stochastic trajectories become jump-like \cite{Young:2018}, and the signal portion of the measurement current, Eq. (\ref{eq:record}), dominates. In this case, $\Pi_{i}(N,t)$ can be estimated as the cumulative probability that the monitored states are populated by the internal dynamics, and stay populated long enough to register a hit.  

In Appendix~\ref{sec:eff_estimation} we present in detail how these probabilities are determined from the average dynamics, which determine the probability of a photo-excitation being transduced into the monitored subsystem, and then the likelihood that the created population will persist long enough to record a hit.  We find that the upper bound for a single detection channel is given by
\begin{flalign}
	\Pi_{i}(t) &= \bar{x}_{i}\dg \bar{\mathcal{G}}(t_{\rm MIN})\bar{x}_{i} \Big[\bar{x}_{i}\dg\bar{\rho} (t-t_{\rm MIN}) - \Big.\nonumber\\
	&~~~~~~~~~~~~~~~~~~~~~~ \Big. (\bar{x}_{i}^{\dagger}\bar{\mathcal{A}}\bar{x}_{i}) \int_{t_0}^{t-t_{\rm MIN}} d\tau~ \bar{x}_{i}\dg\bar{\rho}(\tau)\Big],\label{eq:hitprob}
\end{flalign}
where $t_{\rm MIN}$ is the minimum time for a detection event to be registered. 

If the monitored subsystem is stable and population loss from it can be neglected,  $\bar{x}_{i}^{\dagger}\bar{\mathcal{G}}\bar{x}_{i}=1$ and $\bar{x}_{i}^{\dagger} \bar{\mathcal{A}} \bar{x}_{i}=0$. In this case, the above reduces to 
\begin{flalign}
\Pi_{i}(t)&=\bar{x}_{i} \dg \bar{\rho}(t-t_{\rm MIN}).\label{eq:simplephit}
\end{flalign}
Thus, in this case, the performance can be directly approximated from the average population in the monitored subsystem, consistent with previous results \cite{Young:2018}. 
\subsection{Efficiency}
The total probability that $N$ photons are detected given $M$ incoming photons is $P_N(M,t)=p_N[\Pi_i(M,t)]$, where $p_N$ is a function that maps the outcomes of all output channels into a detection probability. In the case of a single photon $P_1(1,t)=\sum_{i=1}^{n}\Pi_i(1,t)$, where $n$ is the number of detector elements. In that case the mapping function $p_1$ is a simple summation, but for multiple photons it is more complex, as discussed in Appendix~\ref{sec:eff_estimation}.  
In general, the total probability of detecting exactly $N$ photons in a field containing $N$ photons is then $P_N(N,\infty)$; in this case we will often simply write $P_N(N)$.  

\subsection{Dark Counts}
Dark counts for a detector element occur when a hit is recorded in the absence of a field due to total noise exceeding $I_{{\rm HIT},i}$. This noise may include thermal fluctuations of the system (\ie the system has a finite probability of entering the monitored state in the dark) and fundamental noise to the amplification process, as well as noise arising after the amplification due to additional (classical) signal processing and transduction.  Here we consider only the former contributions, both of which are captured in Eq.~\eqref{eq:record}, and ignore the dark counts due to the classical signal processing chain. Since the amplification noise is Gaussian, the dark count rate $r_{i}$ can be obtained straightforwardly from the amplification and integration time $t_m$ using Eq.~\eqref{eq:record} as 
\begin{flalign*}
r_{i}=\frac{\Pi_i(0,t_m+t_0)}{t_m}+\frac{0.5}{t_m}{\rm erfc}\left(2\sqrt{kt_m}\Delta I_{{\rm HIT},i}\right)
\end{flalign*}
where $\Pi_{i}(0,t_m+t_0)$ is the probability of obtaining a hit in time $t_m$ due to noise (when no photons are present in the field), and $\Delta I_{{\rm HIT},i}$ is the difference in signal between hit and non-hit states of the detector.  When amplification noise is the dominating contributor to the dark counts, minimization of the dark count rate requires stronger amplification and/or longer integration times, which in some cases will limit performance. The overall dark count rate for the photodetection system is given by $R_N=p_N(r_i)$ where $p_N$ is the same mapping function as for the efficiency.

\subsection{Jitter and Latency}

The ultimate limits to jitter and latency in a detector are imposed by the temporal spread of an electromagnetic pulse. While these quantities can be simply defined in the case of a single photon pulse, more care is required when defining them for multiphoton pulses. We shall define jitter and latency with respect to a temporal distribution determined by the pulse profile $E(t)$
\begin{flalign*}
f(t)=N\abs{E(t)}^2\left[\int^{t}_{t_0} d\tau\abs{E(\tau)}^2\right]^{N-1},
\end{flalign*}
which we show below to correspond to the behavior of an ideal intensity detector and arrival time it would register for the $Nth$ photon of a multiphoton pulse. 

We compare this quantity to the distribution of detection times for the $N$th photon in an $M$ photon pulse. Since $P_N(M,t)$ gives the cumulative probability of having registered such a hit at time $t$, $\dot{P}_N(M,t)\Delta t$ represents the probability of a hit being obtained in a short interval $\Delta t$ centered at $t$, which, when normalized to the total probability of a hit, yields a distribution of detection times
\begin{flalign*}
g(t)&=\frac{\dot{P}_{N}(M,t)}{P_{N}(M,\infty)}
\end{flalign*}
For example, in the case of a single detection element $i=1$ this can be expressed as
\begin{flalign*}
g(t)&=\frac{\bar{x}_{i}\dg \bar{\mathcal{G}}(t_{\rm MIN})\bar{x}_{i}\bar{x}_{i}\dg\left(\dot{\bar{\rho}} (t-t_{\rm MIN})-\bar{x}_{i}^{\dagger}\bar{\mathcal{A}}\bar{x}_{i}\bar{\rho}(t-t_{\rm MIN})\right)}{\Pi_1(M,\infty)}.
\end{flalign*}

The latency in the detection is then defined as the difference between the mean detection time from this distribution and the mean time from $f(t)$, \ie
\begin{flalign*}
\mu=\int_{t_0}^\infty dt\, t\left[g(t)-f(t)\right].
\end{flalign*}
Similarly, the standard deviation of the detection time for the $N$th photon gives the jitter, \ie
\begin{flalign*}
\sigma&=\sqrt{\int_{t_0}^\infty dt\, t^2g(t)-\left(\int_{t_0}^\infty dt\, tg(t)\right)^2}.
\end{flalign*}
For convenience we define $\sigma_{{\rm SYS}}$ as the jitter originating from the detector so that
\begin{flalign*}
\sigma&=\sqrt{\left(\sigma_0\right)^2+\left(\sigma_{{\rm SYS}}\right)^2}
\end{flalign*}
with
\begin{flalign*}
\sigma_0&=\sqrt{\int_{t_0}^\infty dt\, t^2f(t)-\left(\int_{t_0}^\infty dt\, tf(t)\right)^2}.
\end{flalign*}

\section{Ideal Detection}
The above naturally leads to a definition of ideal detection: a pulse arriving at time $t$ is immediately and fully transduced to a monitored state, such that $g(t)=f(t)$, and thus
\begin{flalign}
P_{N}(N,t)=\left[\int_{-t_0}^{t} d\tau \abs{E(\tau)}^2\right]^N\label{eq:gen_ideal}.
\end{flalign}
Since $\abs{E(\tau)}^2$ is normalized to 1 (see Appendix \ref{sec:lightfield}), we have $P_{N}(N,\infty)=1$ in this case, corresponding to 100$\%$ efficiency.  Additionally, since the distribution of the detection times is equivalent to the temporal distribution of the photon(s) in the pulse, $\mu=0$ and $\sigma=\sigma_0$. Furthermore, one can choose an amplification rate, $k\gg 1/(\chi^2 t_m)$, to make $R_{N} \approx 0$, and achieve dark count rates that are arbitrarily close to zero.       

The above conditions on the metrics can be translated into conditions on the Green's functions governing the dynamics and the underlying architecture. In general, $\bar{\mathcal{G}}$ will have two effects on the signal propagation: one, it may directly attenuate the signal, and two, it may act to alter the shape of the signal that is passed on to the next integral, which will reduce the efficiency.  To obtain Eq.~\eqref{eq:gen_ideal} from Eq.~\eqref{eq:simplephit}, $\bar{\mathcal{G}}$ must act as a delta function when acting on density vectors diagonal in the field, and must be a constant (\ie only comprise modes with zero eigenvalues) when acting on density vectors off-diagonal.  Additionally, the overall magnitude must be unity. This will be shown concretely in the Examples.

\section{Single-Photon, Single Element Detector}
\label{sec:1ph_1det}

\begin{figure}
	\includegraphics[width=\columnwidth]{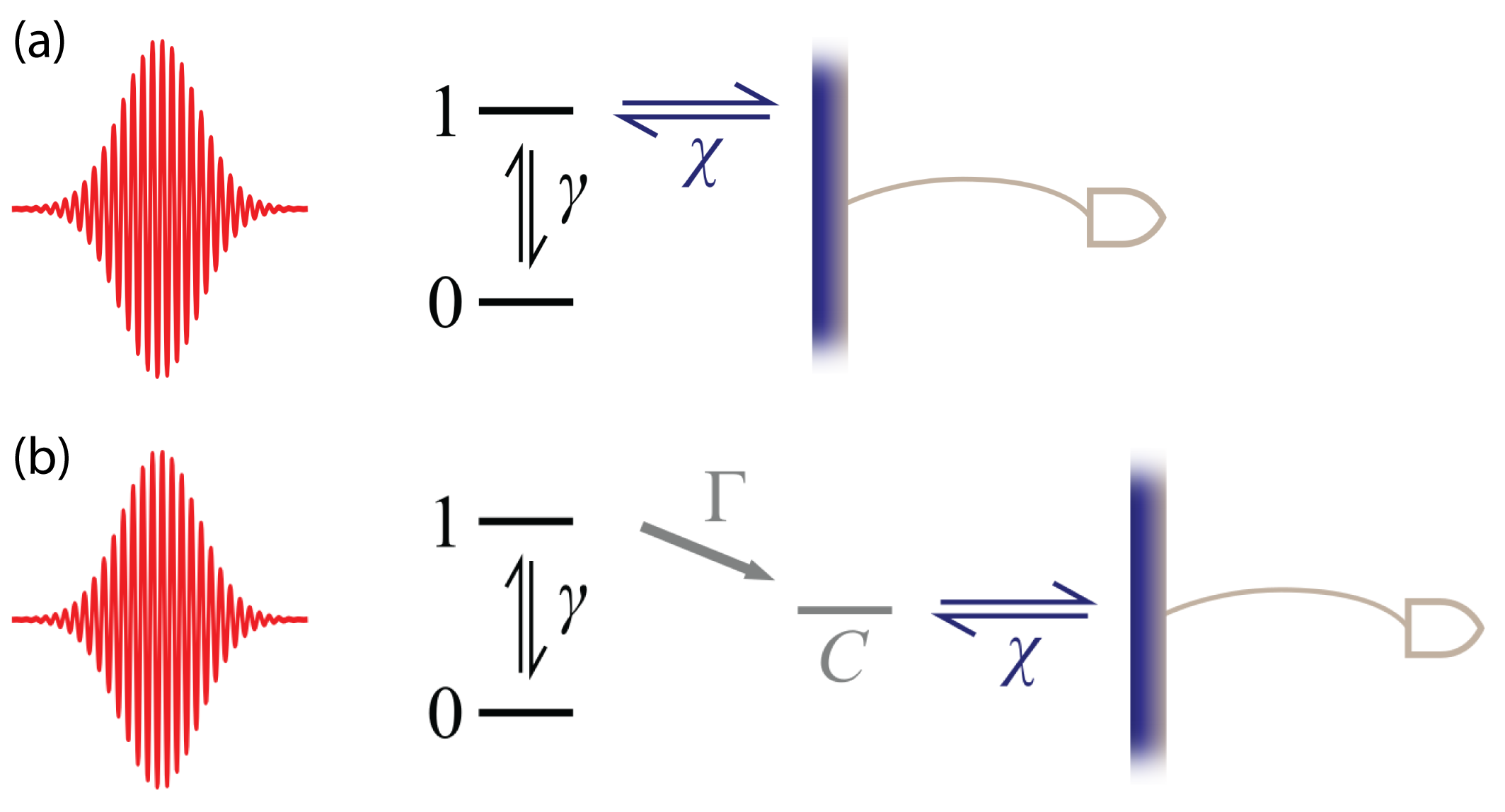}
	\caption{The two- and three-state systems analyzed in Section V.  A single mode couples the ground state 0 and excited state 1 with strength $\gamma$. (a) In the two-state system the 1 state is directly amplified with strength $\chi$ and rate $k$. (b) In the three-state system there is an incoherent decay from state 1 to a state $C$, which is amplified. \label{fig:sys1}}
\end{figure}

To illustrate the Green's functions formalism we have developed, we first consider the simplest model for a detector, a two-state system (see Fig.~\ref{fig:sys1}(a)). A single mode couples states 0 and 1 with strength $\gamma$, while state 1 is amplified with strength $\chi$ and rate $k$.
The matrices governing the dynamics are 

\begin{flalign*}
\hat{L}=\left[\begin{array}{cc}
0 & \gamma \\
0 & 0 
\end{array}\right],
\hat{X}=\left[\begin{array}{cc}
0 & 0\\
0 & \chi
\end{array}\right],
\hat{H}=\left[\begin{array}{ccc}
0 & 0 \\
0 & \omega_{1} \\
\end{array}\right].
\end{flalign*} 

Then, with
\begin{flalign*}
\bar{\rho}&=\left[\begin{array}{c}
\hat{\rho}_{00}\\
\hat{\rho}_{11}\\
\hat{\rho}_{01}\\
\hat{\rho}_{10}
\end{array}\right],
\end{flalign*}

we have 
\begin{flalign*}
\bar{\mathcal{A}}&=\left[\begin{array}{c c c c}
0 & \gamma^2  & 0 & 0\\
0 &-\gamma^2 & 0 & 0 \\
 0 & 0 & -\frac{\gamma^2+2k\chi^2}{2} +i\omega_1 & 0\\ 
 0 & 0 & 0 &-\frac{\gamma^2+2k\chi^2}{2} -i\omega_1 
\end{array}\right],
\end{flalign*}
which can be exponentiated to obtain
\begin{flalign*}
\bar{\mathcal{G}}(t)&=\left[\begin{array}{c c c c}
1 & 1-\expn{-\gamma^2t} & 0 & 0\\
0 & \expn{-\gamma^2t}  & 0 & 0\\
0 & 0 & \expn{\left(i\omega_1-\frac{2k\chi^2+\gamma^2}{2}\right)t} & 0\\
0 & 0 & 0& \expn{\left(-i\omega_1-\frac{2k\chi^2+\gamma^2}{2}\right)t}
\end{array}\right].
\end{flalign*}
The field coupling is described by
\begin{flalign*}
\bar{\mathcal{L}}^+=-\bar{\mathcal{L}}^{-\dagger}=\left[\begin{array}{c c c c}
0 & 0 & 0 & -\gamma\\
 0 & 0 & 0 & \gamma  \\
\gamma & -\gamma & 0 & 0 \\
0 & 0 & 0 & 0 
\end{array}\right].
\end{flalign*}

For an incident resonant single-photon field, the density matrix is given by $\bar{\varrho}^{11}$ and we have
\begin{flalign*}
\bar{\varrho}^{11}(t)&=\int_{-\infty}^t d\tau\bar{\mathcal{G}}(t-\tau)\bar{\mathcal{L}}^+E(\tau)\expn{-i\omega \tau}\\
&\times\int_{-\infty}^\tau d\tau'\bar{\mathcal{G}}(\tau-\tau')\bar{\mathcal{L}}^-E^*(\tau')\expn{i\omega \tau'}\bar{\rho}(t_0)+h.c.,
\end{flalign*}
or explicitly in terms of the population of the excited state,
\begin{flalign*}
\hat{\varrho}^{11}_{11}(t)&=2\gamma^2\int_{-\infty}^t d\tau\expn{-\gamma^2(t-\tau)} E(\tau)\\
&\times\int_{-\infty}^\tau d\tau'\expn{-\frac{2k\chi^2+\gamma^2}{2}(\tau-\tau')}\gamma E^*(\tau').
\end{flalign*}

In order to obtain estimates for the detector metrics we apply the estimation scheme described by Eq.~\eqref{eq:hitprob}.  Using the above definitions for $\bar{\mathcal{G}}(t)$ and $\bar{X}$, 
\begin{flalign}
P_{1}(1,t)&=\expn{-\gamma^2t_{\rm MIN}}\int_{-\infty}^{t-t_{\rm MIN}}d\tau 2\gamma^2E(\tau)\nonumber\\
&\times \int_{-\infty}^\tau d\tau'\expn{-\frac{2k\chi^2+\gamma^2}{2}(\tau-\tau')} E^*(\tau').\label{eq:2Lest}
\end{flalign}

\begin{figure}
	\includegraphics[width=\columnwidth]{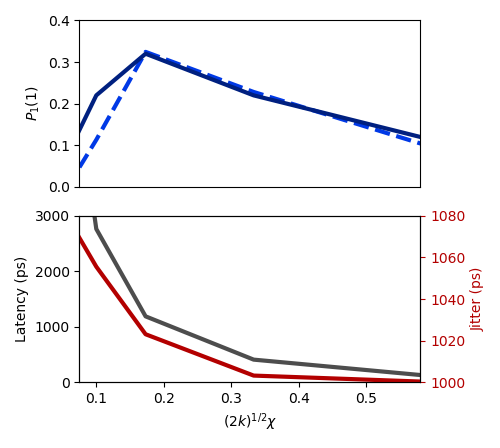}
	\caption{The estimated efficiency vs simulated efficiency (upper panel) and the estimated jitter and latency (lower panel) for the system in Fig.~\ref{fig:sys1}(a).  The simulation results are taken from Ref.~\cite{Young:2018} and the estimation is performed using the same parameters and Eq.~\eqref{eq:2Lest}. The estimated quantity deviates from the calculated quantity in the small $(2k)^{1/2}\chi$ region because the approximations that go into the estimated quantity (\ie jump-like trajectories) are not valid in this very weak measurement regime.  \label{fig:2Lesteff}
}
\end{figure}

With this we can compare against the results obtained in Ref.~\cite{Young:2018} via direct simulation. Using a Gaussian pulse with $\sigma_{E}=1$ns and  $\gamma^2=1$ns$^{-1}$, we compute $P_{1}(1,\infty)$ using the intervals $t-\tau$ corresponding to thresholds that yield the dark count rate in that work (i.e., a probability of 1\% for obtaining a dark count in a 15ns interval) and plot it against the efficiency obtained from direct simulation (Fig.~\ref{fig:2Lesteff}). In this case the estimation scheme provides a good approximation of the efficiency. The estimation deviates more for low $(2k)^{1/2}\chi$,  which is expected: the approximation of the dynamics being jump-like is less valid, the contribution of noise is increased, and the longer integration times increase the likelihood that a direct simulation signal will be split across two integration windows and fail to produce a hit, an event that is not captured by the estimation scheme as presented.  We also plot the estimated jitter and latency, shown in the lower panel.  These are reduced for stronger amplification consistent with the trends observed in Ref.~\cite{Young:2018}.  

We next consider the system containing three states: 0,1, and $C$ (Fig.~\ref{fig:sys1}(b)). A single, resonant mode couples 0 and 1 with strength $\gamma$, while state 1 decays incoherently to state $C$ according to $\Gamma$, which is amplified according to $\chi$ and $k$.  We previously demonstrated that this configuration can lead to ideal detection with 100\% efficiency, negligible dark counts, and minimal jitter, provided that certain conditions are met for $\gamma$ and $\Gamma$ \cite{Young:2018}.

For this system we have
\begin{flalign*}
\hat{L}=\left[\begin{array}{ccc}
0 & \gamma & 0\\
0 & 0 & 0\\
0 & 0 & 0
\end{array}\right],
\hat{Y}=\left[\begin{array}{ccc}
0 & 0 & 0\\
0 & 0 & 0\\
0 & \Gamma & 0
\end{array}\right],\\
\hat{X}=\left[\begin{array}{ccc}
0 & 0 & 0\\
0 & 0 & 0\\
0 & 0 & \chi
\end{array}\right],
\hat{H}=\left[\begin{array}{ccc}
0 & 0 & 0\\
0 & \omega_{1} & 0\\
0 & 0 & \omega_{C}
\end{array}\right].
\end{flalign*} 
We construct the vectorized density matrix as
\begin{flalign*}
\bar{\rho}=\left[\begin{array}{c}
\hat{\rho}_{00}\\
\hat{\rho}_{11}\\
\hat{\rho}_{CC}\\
\hat{\rho}_{01}\\
\hat{\rho}_{10}
\end{array}\right].
\end{flalign*}
Coherences with the $C$ state are omitted as they will always be zero for the given interactions and initial state $\hat{\rho}(t_0)=\ket{0}\bra{0}$.  
Then 
\begin{flalign*}
\bar{\mathcal{A}}&=\left[\begin{array}{c c c c c}
0 & \gamma^2 & 0 & 0 & 0 \\
0 &  -(\gamma^2+\Gamma^2) & 0 & 0 & 0 \\
0 &  \Gamma^2 & 0 & 0 & 0\\
0 & 0 & 0 & -\frac{\gamma^2+\Gamma^2}{2} +i\omega_1^2 & 0 \\
0 & 0 & 0 & 0 & -\frac{\gamma^2+\Gamma^2}{2} -i\omega_1^2 
\end{array}\right]
\end{flalign*}
and
\begin{flalign*}
\bar{\mathcal{L}}^+=-\bar{\mathcal{L}}^{-\dagger}=\left[\begin{array}{c c c c c}
0 & 0 & 0 & 0 & -\gamma\\
0 & 0 & 0 & 0 & \gamma  \\
0 & 0 & 0 & 0 & 0 \\
\gamma & -\gamma & 0 & 0 & 0 \\
0 & 0 & 0 & 0 & 0
\end{array}\right].
\end{flalign*}
This leads to
\begin{flalign*}
\bar{\mathcal{G}}(t)&=\left[\begin{array}{c c c c c}
1 & \gamma^2\frac{1-\expn{-(\Gamma^2+\gamma^2)t}}{\Gamma^2+\gamma^2} & 0 & 0 & 0\\
0 & \expn{-(\Gamma^2+\gamma^2)t} & 0 & 0 & 0\\
0 & \Gamma^2\frac{1-\expn{-(\Gamma^2+\gamma^2)t}}{\Gamma^2+\gamma^2} & 1 & 0 & 0\\
0 & 0 & 0 & \expn{\left(i\omega_1-\frac{\Gamma^2+\gamma^2}{2}\right)t}  & 0 \\
0 & 0 & 0 & 0 & \expn{\left(i\omega_1-\frac{\Gamma^2+\gamma^2}{2}\right)t}  
\end{array}\right].\\
\end{flalign*}

We note two features of the above matrices: first, the top $3\times3$ diagonal block of $\bar{\mathcal{A}}$ only has a single nonzero eigenvalue, a decay mode, and second, none of these matrices have a dependence on the overall amplification $(2k)^{1/2}\chi$. The separation of the unmonitored and monitored subspaces by an incoherent decay, and the fact that there are only incoherent processes within the monitored subspace implies that the amplification does not directly effect the dynamics.  

For a single photon at the resonant field frequency and a system initially in state 0, we find that the probability of a hit, which is equal to the population in the $C$ state in this case, is 

\begin{flalign}
P_1(1,t)&=\hat{\varrho}^{11}_{CC}(t) =2\frac{\Gamma^2\gamma^2}{\gamma^2+\Gamma^2}\nonumber\\
&\times\int_{-\infty}^td\tau\left(1-\expn{-(\gamma^2+\Gamma^2)(t-t')}\right) E(\tau)\nonumber\\
&\hspace{2cm}\times\int_{-\infty}^\tau d\tau'\expn{-\left(\frac{\gamma^2+\Gamma^2}{2}\right)(\tau-\tau')}E^*(\tau').\label{eq:sphmd}
\end{flalign}
When $\gamma^2+\Gamma^2 \gg 1/\sigma_E$, where $\sigma_E$ is the width of a smooth pulse described by $E(t)$, the exponential decay can be approximated by a delta function.  Then
\begin{flalign*}
P_1(1,t)=2&\frac{\gamma^2\Gamma^2}{\gamma^2+\Gamma^2}
\int_{-\infty}^td\tau \abs{E(\tau)}^2\frac{2}{\gamma^2+\Gamma^2},\\
P_1(1)=&4\frac{\gamma^2\Gamma^2}{(\gamma^2+\Gamma^2)^2}.
\end{flalign*}
This expression satisfies Eq.~\eqref{eq:gen_ideal} except for the prefactor.  It is evident by inspection that this prefactor is unity when $\gamma=\Gamma$.  Thus, Eq.~\eqref{eq:gen_ideal} can be fully satisfied and perfect efficiency is achieved when the two rates are equal; the large coupling limit ($\gamma^2+\Gamma^2 \gg 1/\sigma_E$) ensures that the signal is not distorted as it is processed by the system, while the coupling matching condition ($\gamma=\Gamma$) ensures the photon is fully converted to population in $C$.  We emphasize that as long as the pulse is sufficiently long, this result is obtained regardless of pulse shape.  

We also note that the fast incoherent process (large $\Gamma$) will widen the detection bandwidth and decrease sensitivity to the resonance condition.  Repeating the above calculation for an off-resonant pulse we obtain
\begin{flalign*}
P_1(1,t)= 2&\frac{\Gamma^2}{\gamma^2+\Gamma^2}\int_{-\infty}^td\tau E(\tau)
 \int_{-\infty}^\tau d\tau'\gamma^2\nonumber\\
 &\times\cos \left[\Delta\omega(\tau-\tau')\right]\expn{-\left(\frac{\gamma^2+\Gamma^2}{2}\right)(\tau-\tau')}E^*(\tau'),\\
P_1(1)=&4\frac{\gamma^2\Gamma^2}{(\gamma^2+\Gamma^2)^2+4\Delta\omega^2},
\end{flalign*}
where $\Delta\omega$ is the detuning from resonance. The bandwidth of the detector is therefore determined by the rate $\gamma^2+\Gamma^2$. The requirement that the pulse be temporally wide can also be understood as requiring that the pulse frequency distribution be narrow compared to this bandwidth, such that it is approximately constant over the frequencies in the pulse.  

It is also evident from this calculation that any additional sources of decoherence, such as amplification or decays to unmonitored states, will reduce the efficiency; they will appear in the denominator of the prefactor and the maximum efficiency will no longer be unity. For example, suppose that we have additional dephasing of the excited state, modeled by the Lindblad operator $\hat{Y}_1=\kappa\ket{1}\bra{1}$.  This will inhibit the formation of optical coherence but will not affect the populations.  Adding this dynamics gives
\begin{flalign*}
P_1(1,t)=2&\frac{\Gamma^2}{\gamma^2+\Gamma^2}\int_{-\infty}^td\tau E(\tau)
\int_{-\infty}^\tau d\tau'\gamma^2\nonumber\\
&\times\expn{-\left(\frac{\gamma^2+\Gamma^2+\kappa^2}{2}\right)(\tau-\tau')}E^*(\tau'),\\
P_1(1)=&4\frac{\gamma^2\Gamma^2}{(\gamma^2+\Gamma^2)^2}\frac{\gamma^2+\Gamma^2}{\gamma^2+\Gamma^2+\kappa^2}.
\end{flalign*} 
These expressions indicate that the efficiency is the efficiency obtained without the dephasing dynamics multiplied by an additional factor dependent on $\kappa$.  This factor --  and the total efficiency -- will always be less than 1.

Our formalism also reveals the critical role of the zero eigenvalue mode of $\bar{\mathcal{A}}$ in the dynamics; the presence of a zero eigenvalue indicates a component of the population that is persistent; the state will be populated indefinitely.  Decays or oscillations would prohibit an expression in the form of Eq.~\eqref{eq:gen_ideal} from being achieved. 
For example, suppose that population decays from the $C$ state to the $0$ state at a rate $\delta^2$, so that the detector effectively resets after time $\sim 1/\delta^{2}$.  Then, even with $\gamma=\Gamma$,
\begin{flalign}
P_1(1,t)&=\int_{-\infty}^td\tau\expn{-\delta^{2}t_{\rm MIN}}\abs{E(t-t_{\rm MIN})}^2,\nonumber\\
P_1(1)&=\expn{-\delta^{2}t_{\rm MIN}},\label{eq:reseteff}
\end{flalign}
so that the overall efficiency is limited by the possibility that population in $C$ decays before sufficient time to record a hit has passed. Maximizing efficiency requires both separation of the amplification from the absorption and propagation of the signal into such stationary system mode where it can be amplified at leisure, i.e., shelving. However, this extends the natural reset time, limiting count rate in the absence of an active reset mechanism. 

In practice the time required for the system to record a hit places additional constraints on the amplification $(2k)^{1/2}\chi$.  The maximum threshold that can produce a hit is $\Delta I_{{\rm HIT}}=\chi$.  For a desired dark count rate $R_1$, then $(2k)^{1/2}\chi\ge\frac{{\rm erfc}^{-1}\left(2t_m R_1\right)}{\sqrt{2t_m}}$.  This is especially relevant if relaxation from the $C$ state to the ground state is present.  Maximizing the efficiency in Eq.~\eqref{eq:reseteff} requires increasing the amplification so that $t_m$ may be reduced, or reducing the signal threshold and necessarily permitting more dark counts. 

\section{Other detector structures}
\label{sec:other}

To demonstrate the broad applicability of the formalism, we now apply it to several cases with other
detection mechanisms or detector configurations.

\subsection{Two-State Quadratic Coupling Detector}
In some cases the matter system couples to the field quadratically; the interaction does not alter the photon number but the phase of the field. For example, in circuit-QED architectures in the dispersive regime, the interaction between matter and field takes the form $H_{I} = a\dg a \hat{\sigma}_z$ \cite{blais_cavity_2004}. In this case, the natural way to model the field-matter interaction within our framework is to couple through $\hat{S}$ rather than $\hat{L}$ in Eq.~(\ref{eq:full_me}).  Instead of creating optical coherence, the field scatters off the matter subsystem and causes a change. Consider a two-level system such as that in Fig.~\ref{fig:sys1}(a), with the field scattering causing a general unitary operation on the matter degrees of freedom.  Then the system is described by the operators 
\begin{flalign*}
\hat{S}=\expn{-i\left(a_x\hat{\sigma}_x+a_y\hat{\sigma}_y+a_z\hat{\sigma}_z\right)\theta/2},
\hat{H}=\left[\begin{array}{cc}
0 &0\\
0 &\omega_1
\end{array}\right],
\hat{X}=\left[\begin{array}{cc}
0 &0\\
0 &\chi
\end{array}\right],
\end{flalign*}
where $\hat{\sigma}_x,\hat{\sigma}_y,\hat{\sigma}_z$ are Pauli matrices and $a_x^2+a_y^2+a_z^2=1$.
This gives 
\begin{flalign*}
\bar{\mathcal{G}}(t)&=\left[\begin{array}{cccc}
1 &0&0&0\\
0&\expn{i\omega_1 t}&0&0\\
0&0&\expn{i\omega_1 t}&0\\
0&0&0&1
\end{array}\right],\\
\bar{\mathcal{S}}&=\sin^2(\theta/2)\left(\hat{\alpha}\otimes\hat{\alpha}^{\rm T}-\bv{I}\right)\\
&+i\sin(\theta/2)\cos(\theta/2)\left(\hat{\alpha}\otimes\bv{I}-\bv{I}\otimes\hat{\alpha}^{\rm T}\right),
\end{flalign*} 
where $\hat{\alpha}=\left(a_x\hat{\sigma}_x+a_y\hat{\sigma}_y+a_z\hat{\sigma}_z\right)$.
This yields,
\begin{flalign*}
P_1(1,t)=\hat{\varrho}^{11}_{11}=\frac
{\sin^2\theta}{2(1-a_z^2)
}\int_{-\infty}^{t-t_{\rm MIN}} d\tau\abs{E(\tau)}^2,
\end{flalign*}
so ideal detection is achieved when $a_z = 0$ and $\theta=\pi$.
The quadratic field coupling does not produce a field-driven decay, and can directly transfer population.  As a consequence there are no losses and the requirement for ideal detection is straightforwardly satisfied without requiring a shelving state.  Even when ideal efficiency is not obtained, the quadratic coupling ensures that $\dot{P}_1(1,t)=\abs{E(t-t_{\rm MIN})}^2$, meaning jitter is minimized and latency is limited by the integration time of amplification.       

\subsection{Single Photon, Multiple Degenerate Elements}
\label{sec:1ph_mdet_degen}
\begin{figure}
	\includegraphics[width=\columnwidth]{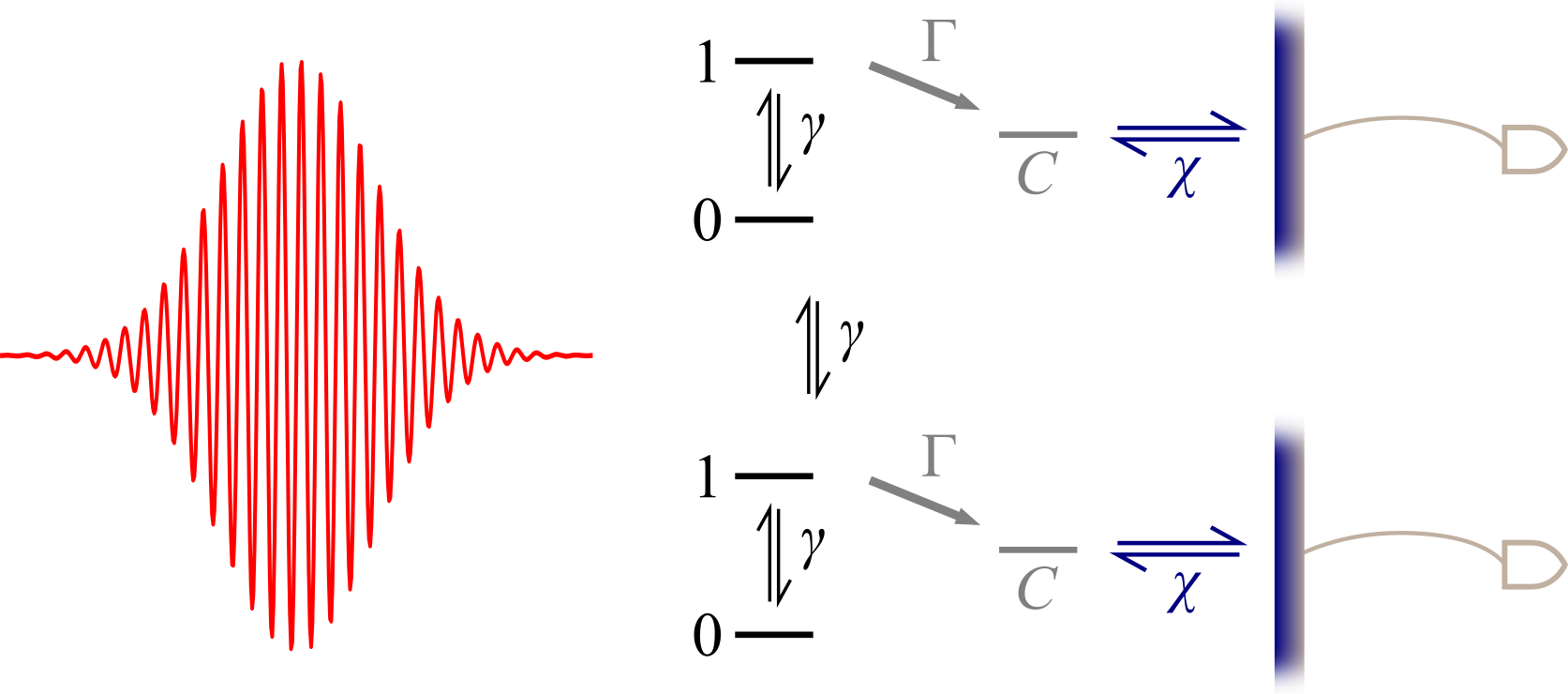}
	\caption{System comprising two elements described by Fig.~\ref{fig:sys1}(b). The elements collectively interact with the field and may exchange population through the field. In systems with more elements, the field couples each element with each other element. \label{fig:sys2x1}}
\end{figure}
We next consider $n$ copies of the elements of Fig.~\ref{fig:sys1}(b) collectively interacting with the incoming light pulse, illustrated for two elements in Fig.~\ref{fig:sys2x1}. Each absorbing element is assumed to be degenerate in the sense that all elements have the same transition energy between states $0$ and $1$, and each absorbing element has its own associated readout channel. The system Hilbert space is the direct product of $n$ three-state Hilbert spaces.  
There is a single operator coupling the system with the field and a bath coupled to each subsystem such that
\begin{flalign*}
\hat{L}=\sum_i\gamma\ket{0_i}\bra{1_i}, \hat{Y}_i=\Gamma\ket{C_i}\bra{1_i}
\end{flalign*}
where $1_i$ and $C_i$ denote the 1 and $C$ states of the $i$th element.  
As a result the field introduces coupling between array elements whereas the baths $\hat{Y}_i$ do not; following the procedure introduced in Section II  we find that

\begin{flalign*}
P_1(1,t)&=\sum_{i=1}^{n}\Pi_i(1,t)=\sum_{i}\bar{x}_i\cdot\bar{\varrho}^{11}(t) \\
&=2n\frac{\gamma^2\Gamma^2}{n\gamma^2+\Gamma^2}\int_{-\infty}^td\tau\left(1-\expn{-(n\gamma^2+\Gamma^2)(t-t')}\right) \\
&~~~\times E(\tau)\int_{-\infty}^\tau d\tau'\expn{\left(\frac{n\gamma^2+\Gamma^2}{2}\right)(\tau-\tau')}E^*(\tau').
\end{flalign*}
This is the same expression as for the single-element detector with $\gamma\rightarrow\sqrt{n}\gamma$, indicating that a multiple copy detector can be made equivalent to a single copy detector with stronger field coupling, and the ideal detector is now characterized by $\Gamma=\sqrt{n}\gamma$. The presence of multiple degenerate elements enhances the effective field coupling, a phenomenon recognized as superradiance \cite{gross_superradiance:_1982}. 

It is important to note that the need to monitor multiple states may introduce practical constraints on detector performance.  In particular there are three main amplification schemes that are consistent with the above description: i) a single amplification process sensitive to all the $C$ states, corresponding to a single $\hat{X}=\chi\sum_i^n\ket{C_i}\bra{C_i}$, ii) an amplification process monitoring each $C$ state, so that there are $n$ separate $\hat{X}_i=\chi\ket{C_i}\bra{C_i}$, all of which are processed to determine hits individually, and iii) amplification processes for each $C$ state, the combined signal of which is processed to determine whether a hit has occurred.  For a single photon, both i and ii result in the same dark count rate for given amplifications and times, since the hits are determined based on the noise provided by a single amplification process.  However, for iii, the noise is the total contributed by all amplification processes and will be larger than the single-process noise by a factor of $\sqrt{n}$, increasing the difficulty of discriminating hits and dark counts.  

\subsection{Single Photon, Dispersive Band }
\label{sec:1ph_mdet_nondegen}
In most cases the detector will contain multiple absorbing elements that are not degenerate, resulting in a band of possible excitations from the ground state. This is the case in ensembles of atoms or solid state systems.  It is therefore essential to understand how the above results can be extended to such systems.  

\begin{figure}
	\includegraphics[width=\columnwidth]{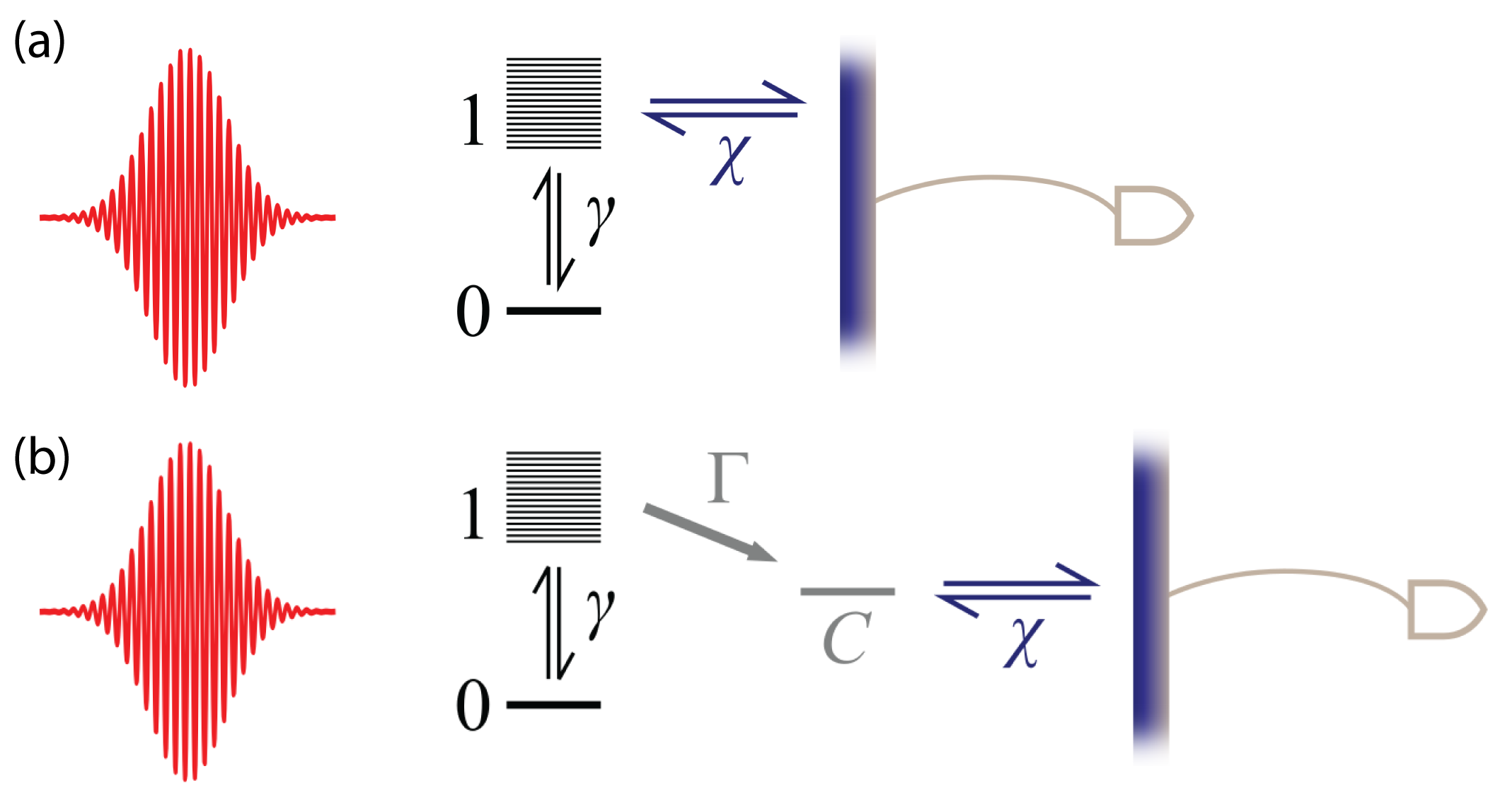}
	\caption{Systems featuring excitation into a dispersive band.   A single mode couples the ground state 0 and all excited states with strength $\gamma$. In (a) the states in the band are directly amplified with strength $\chi$ and rate $k$, while in (b) the excited band states incoherently decay to a state (or potentially multiple states) $C$, which are amplified. \label{fig:sys_band}}
\end{figure}

We consider systems like those appearing in Fig.~\ref{fig:sys1}, which have a single ground state that may be excited with strength $\gamma$ to $n$ independent states with energies $\omega$ distributed according to a normalized function $D(\omega)$.
First, we consider the case with no amplification dynamics in order to isolate the effect of having a band of states that are coupled to  the field.

In the limit $n\rightarrow \infty$, we treat the distribution of excited states as a continuous density, and in the case where the distribution of these states is Lorentzian we can analytically compute this excited state population. For a Lorentzian with full-width-half-max $\zeta^2/2$ and centered at the incoming pulse frequency $\omega_E$, the state distribution is
\begin{flalign*}
D(\omega)=\frac{1}{\pi}\frac{\zeta^2/2}{(\zeta^2/2)^2+(\omega-\omega_E)^2},
\end{flalign*}
and the total population of the band is
\begin{flalign*}
u(t)&=\sum_i\bar{\varrho}^{11}_{ii}(t)=\frac{ 2\gamma^2}{n\gamma^2+\zeta^2}\\
&\times\int^t d\tau \left(n\gamma^2\expn{-(n\gamma^2+\zeta^2)(t-\tau)}+\zeta^2\right) E(\tau)\\
&\times\int^\tau d\tau'  \expn{-\frac{n\gamma^2+\zeta^2}{2}(\tau-\tau')} E^*(\tau')
\end{flalign*}
where $i$ indexes the states in the band.  

The form of this expression is that of Eq.~\eqref{eq:sphmd} and the Green's functions have the required form for ideal detection. The energy dispersion results in dephasing that suppresses emission allowing the excitation into the band to be long lived.   From the above expression it is clear that the bandwidth can be tuned to obtain full collection of the photon when $\zeta^2=n\gamma^2 \gg \sigma_0$.  However, this does not necessarily indicate ideal performance; unlike the case of incoherent decay to a shelving state, amplification will influence the absorption process with consequences for performance. If the excited states are directly amplified according to $\hat{X}=\chi\sum^n_i\ket{i}\bra{i}$ with rate $k$, then we can write
\begin{flalign*}
P_1(1,t)&=u(t)=\frac{ 2\gamma^2}{n\gamma^2+\zeta^2}\\
&\times\int^t d\tau \left(n\gamma^2\expn{-(n\gamma^2+\zeta^2)(t-\tau)}+\zeta^2\right) E(\tau)\\
&\times\int^\tau d\tau'  \expn{-\frac{n\gamma^2+\zeta^2+2k\chi^2}{2}(\tau-\tau')} E^*(\tau')
\end{flalign*}
and in the limit $\zeta^2=n\gamma^2 \gg \sigma_0$
\begin{flalign*}
P_1(1,t)=\frac{ 4\gamma^2\zeta^2}{(n\gamma^2+\zeta^2)^2}\times\frac{n\gamma^2+\zeta^2}{n\gamma^2+\zeta^2+2k\chi^2}.
\end{flalign*} 

The amplification introduces decoherence that suppresses absorption and will limit efficiency; \ie~the Zeno effect, see 
Ref.~\cite{Young:2018}. However, unlike the two-state system, the band shelving-like effect allows for long measurement intervals and correspondingly lower amplifications; this suggests that higher efficiencies may be obtained at the expense of increased latency. 

We may circumvent such tradeoffs as before by introducing incoherent decays from each state in the band to states that are not optically coupled (dark states) that are then amplified.  It is straightforward to alter the above treatment to account for the additional dynamics. For a decay rate $\Gamma^2$ to a single state $C$ (or multiple $C$ states),  the band population becomes
\begin{flalign*}
u(t)&=\int^t d\tau \expn{-\Gamma^2(t-\tau)}\frac{n\gamma^2\expn{-(n\gamma^2+\zeta^2)(t-\tau)}+\zeta^2}{n\gamma^2+\zeta^2}\\
&\times 2\gamma E(\tau)\int^\tau d\tau'  \expn{-\frac{n\gamma^2+\zeta^2+\Gamma^2}{2}(\tau-\tau')}\gamma E^*(\tau'),
\end{flalign*}
and 
\begin{flalign*}
P_1(1,t)&=\bar{\varrho}_{CC}(t)=\Gamma^2\int^t d\tau p(\tau)\\
&=\int^t d\tau \frac{\Gamma^2}{n\gamma^2+\zeta^2}\left[\frac{n\gamma^2(1-\expn{-(n\gamma^2+\zeta^2+\Gamma^2)(t-\tau)})}{n\gamma^2+\zeta^2+\Gamma^2}\right.\\
&\left.+\frac{\zeta^2(1-\expn{-\Gamma^2(t-\tau)})}{\Gamma^2}\right]\\
&\times 2\gamma E(\tau)\int^\tau d\tau'  \expn{-\frac{n\gamma^2+\zeta^2+\Gamma^2}{2}(\tau-\tau')}\gamma E^*(\tau').
\end{flalign*}
When $\Gamma^2+\zeta^2+n\gamma^2 \gg \sigma_0$, after the pulse passes
\begin{flalign*}
P_1(1)=&4\frac{n\gamma^2(\Gamma^2+\zeta^2)}{(n\gamma^2+\zeta^2+\Gamma^2)^2}
\end{flalign*}
which is unity when $\Gamma^2+\zeta^2=n\gamma^2$. Thus, in the presence of a band, the optimal tuning of the decay rate $\Gamma^2$ has contributions from the bandwidth as well as the optical coupling. However, the jitter and latency both may still be impacted. For the above parameters, 
\begin{flalign*}
P_1(1,t)&=\int^t d\tau\left[ 1- \frac{\expn{-\Gamma^2(t-\tau)}}{2+\Gamma^2/\zeta^2}\right]\abs{E(\tau)}^2.
\end{flalign*}
If $\Gamma$ is small, the second term in brackets is only zero at long times, which delays and stretches the distribution of arrival times.  This adds latency
\begin{flalign*}
\mu=\frac{1}{\Gamma^2\left(2+\Gamma^2/\zeta^2\right)}
\end{flalign*}
and increases the jitter by
\begin{flalign*}
&\sigma_{\rm SYS}=\mu\sqrt{3+2\Gamma^2/\zeta^2}.
\end{flalign*}

\subsection{Multiple Photons, Multi-element Detector}
\label{sec:mph_mdet}

We can now ask how an ideal single-photon detector performs in resolving photon number. For the array of degenerate elements, we note that in the limit where $n\gamma^2+\Gamma^2\gg1/\sigma_0$, the populations of the 1 states and coherences between them are close to 0.   Thus for two photons we have 
\begin{flalign*}
\bar{\varrho}^{22}(t)\approx&\left[1-ny_n(t)\right]\bar{\rho}^{11}_{n:00}+\sum_i^ny_n(t)\bar{\rho}^{11}_{n:C_iC_i}
\end{flalign*}
where
\begin{flalign*}
y_n(t)&=\frac{4\gamma^2\Gamma^2}{\left(n\gamma^2+\Gamma^2\right)^2}\int_{-\infty}^td\tau \abs{E(\tau)}^2. 
\end{flalign*}
Here
$\bar{\rho}^{11}_{n:00}$ and $\bar{\rho}^{11}_{n:C_iC_i}$ represent the total system density matrix in a compact
notation, see Appendix \ref{sec:degenerate} for details.
We can express $\bar{\varrho}^{11}(t)$ as
\begin{flalign*}
\bar{\varrho}^{11}_n(t)=\bar{\varrho}^{00}_{n-1}(t)\otimes\left[\begin{array}{c}
1-y_n(t)\\
0\\
y_n(t)\\
0\\
0
\end{array}\right]=\bar{\rho}_{n-1}(t_0)\otimes\left[\begin{array}{c}
1-y_n(t)\\
0\\
y_n(t)\\
0\\
0
\end{array}\right],
\end{flalign*}
and thus for a system illuminated by a two photon pulse
\begin{widetext}
\begin{flalign*}
\bar{\varrho}^{22}_n&=\bar{\rho}_n(t_0)-\left[\int_{-\infty}^td\tau\expn{\bar{\mathcal{A}}(t-\tau)}\sqrt{2}E(\tau)\expn{-i\omega\tau }\bar{\mathcal{L}}\int_{-\infty}^\tau d\tau'\expn{\bar{\mathcal{A}}(\tau-\tau')}\sqrt{2}E^*(\tau')\expn{i\omega\tau'}\bar{\mathcal{L}}^\dagger\bar{\varrho}_n^{11}(\tau')+h.c.\right]\nonumber\\
\end{flalign*}
We note that in order to determine photon number it is necessary for the amplification scheme to distinguish the states with different numbers of $C$ states occupied.  At present we take each $C_i$ state to be amplified by an operator $\hat{X}_i=\chi\ket{C_i}\bra{C_i}$, so that, as discussed in Appendix~\ref{sec:eff_estimation},
\begin{flalign}
P_2(2,t)=\sum^n_{i> j}\left(\bar{x}_i\times\bar{x}_j\right)\cdot \bar{\varrho}^{22}_n(t)
&=\frac{8\gamma^2\Gamma^2}{\left((n-1)\gamma^2+\Gamma^2\right)^2}\frac{4\gamma^2\Gamma^2}{\left(n\gamma^2+\Gamma^2\right)^2}\frac{\left[\int_{-\infty}^td\tau \abs{E(\tau)}^2\right]^2}{2}\label{eq:mphot}
\end{flalign}
\end{widetext}
where $\times$ represents element by element multiplication.

\begin{figure}
	\includegraphics[width=\columnwidth]{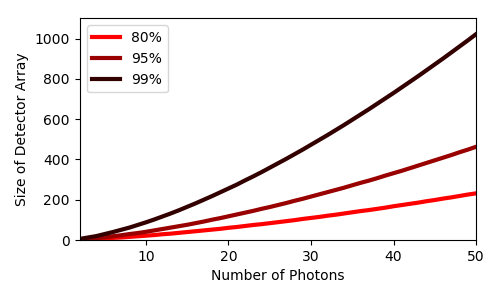}
	\caption{Number of elements in a detector array required to detect a number of photons with a certain efficiency. The detector elements have the configuration in Fig.~\ref{fig:sys2x1}.}\label{fig:array_eff}
\end{figure}

Repeating this procedure, we may find the population of the state corresponding to collection of all $N$ photons at any time $t$. Accounting for the degeneracy of this state via a binomial coefficient, we obtain the following expression for the probability of complete absorption of all incoming photons:
\begin{flalign}
P_{N}(N)=\binom{n}{N}\prod_{k=0}^{N-1}\frac{4n(k+1)}{(2n-k)^2}\label{eq:Narray1}.
\end{flalign}
It is evident that this detector is only a perfect \emph{single} photon detector; multiple photons are detected increasingly inefficiently as the number of photons increases.  This may be understood by thinking about the absorption of photons as a successive process.  Absorption of the first photon proceeds with maximal efficiency, filling state $C$.  However, the occupation of this state blocks further collection by that state: if an element of the array absorbs a photon, it is no longer available to absorb further photons.  This is seen in Eq.~\eqref{eq:mphot} in the appearance of $n-1$ instead of $n$ multiplying $\gamma^2$.  Thus, the detector interacting with successive photons is increasingly detuned from the optimal ratio, as the effective field coupling is reduced, but $\Gamma$ is not. 
This suggests that high efficiency photon resolution requires an array of many more elements than potential photons detected.  In Fig.~\ref{fig:array_eff}, the number of detector elements required to detect all photons in a pulse with efficiencies of 80\%, 95\%, and 99\% is plotted against the number of photons in the pulse.  As shown, the minimum number of detector elements increases dramatically as more stringent requirements for efficiency are imposed.  

This reveals an important design consideration for high-performance number-resolving detectors: the effective parameters of the detector will be altered by interaction with the photon packet, and an ideal detector architecture is one that continues to be optimally tuned for interaction with the remainder of the wavepacket.  For example, to avoid the need for excessive numbers of elements, the array structure analyzed above must be reconfigured so that either the absorption properties are unchanged as the system absorbs photons, or the decay process is also modified so that the effective decay and effective optical coupling remain matched following absorption.      

\section{Conclusion}

We developed a general framework, valid for any point-like detector system under weak light coupling, for simulating, analyzing, and engineering one and few photon detection.  Based on theoretical treatments of light-matter interactions and open quantum systems, we constructed a unified approach to modeling photodection, deriving an expression that yields an output signal generated through the propagation of the incoming photon occupation by a Green's function representing the matter system.  This expression provides a means of both explicitly calculating detection trajectories for determination of performance metrics as well as efficiently estimating these metrics from averaged solutions.  It naturally furnishes a definition of ideal detection that places constraints on the system Green's function and ultimately the detector structure.  Using this we analytically analyze detector schemes and show that they conform to this definition with appropriate parameters. We also generalize the description of an ideal detector to systems with multiple excited levels, with and without energy dispersion, as well as multiple photon resolution.  

We emphasize that this framework describes detection abstractly and may encompass detectors utilizing very different physical systems, allowing comparative analysis of different detector architectures in a unified way.  
While we find, in principle, that ideal detector performance is possible, the structure of energy levels and couplings between them or the physically allowed parameters may prohibit the realization of this limit.  By representing a detector implementation within the framework, it may be evaluated against the ideal limit, clarifying the constraints and tradeoffs imposed by the system physics.  This can guide engineering of the system architecture and parameters to improve the detector performance as well as facilitate the design of entirely novel detectors. 

\begin{acknowledgments}
	Work supported by the DARPA DETECT program. Sandia National Laboratories is a
	multimission laboratory managed and operated by National Technology and
	Engineering Solutions of Sandia, LLC., a wholly owned subsidiary of Honeywell
	International, Inc., for the U.S. Department of Energy's National Nuclear
	Security Administration under contract DE-NA-0003525. The views expressed in the article do not necessarily represent the views of the U.S. Department of Energy or the United States Government.
\end{acknowledgments}

\appendix
\section{Light Field Construction}
\label{sec:lightfield}
We use fields constructed using the quantum noise formalism where the field annihilation and creation operators $\hat{b}(t)$ and $\hat{b}^\dg(t)$ satisfy the commutation relation $\left[\hat{b}^\dg(t),\hat{b}(t')\right]=\delta(t-t')$.  From this one may define the objects
\begin{flalign}
dB_{t}&=\int_t^{t+dt}d\tau \hat{b}(\tau),\nonumber\\
d\Lambda_{t}&=\int_t^{t+dt}d\tau \hat{b}^\dagger(\tau)\hat{b}(\tau),\nonumber\\
\ket{N}&=\frac{1}{\sqrt{N!}}\left[\int d\tau E(\tau)\hat{b}(\tau)^\dagger\right]^N\ket{0},\nonumber\\
dB_{t}\ket{N}&=dt\sqrt{N}E\ket{N-1},\label{eq:fielddef}
\end{flalign} 
where  $N$ is the number of photons in the $i$th mode with temporal field profile $E(t)$ defined so that $\int_{-\infty}^{\infty} dt\abs{E(t)}^2=1$.  Then the light field is in general $\hat{\rho}_{\rm LIGHT}(t) = \sum_{N,M} c_{N,M}(t)\ket{N}\bra{M}$.

In this manuscript we consider only a single mode; however, the consideration of multiple arbitrary modes is straightforward, if cumbersome.  In this case, each mode will be associated with an index $i$ denoting the mode temporal profile $E_i(t)$ and occupation $N_i$.  The component field states are written as $\ket{\bvN}=\prod_i\ket{N_i}$, where $\bvN$ is a vector containing the $N_i$, so that the overall field density matrix is $\hat{\rho}_{\rm LIGHT}(t) = \sum_{\bvN,\bvM} c_{\bvN,\bvM}(t)\ket{\bvN},\bra{\bvM}$.  As a result, the matter state is $\hat{\rho}_{\rm MATTER}(t)=c_{\bvN,\bvM}(t_0)\hat{\varrho}^{\bvN,\bvM}(t)$ and the $\hat{\varrho}^{\bvN\bvM}(t)$ are dependent on all $\hat{\varrho}^{\bvN-1_i,\bvM}(t)$, $\hat{\varrho}^{\bvN,\bvM-1_i}(t)$, and $\hat{\varrho}^{\bvN-1_i,\bvM-1_i}(t)$ where $\bvN-1_i$ is the overall occupation $\bvN$ less one photon in the $i$th mode.

\section{Connection to POVM}
\label{sec:povm}
Any quantum measurement can be mathematically represented as a positive operator-valued measure (POVM) \cite{kraus_states_1983,nielsen_quantum_2010}, which is a set of operators/elements defined on the Hilbert space of the system being measured. Each of these operators is labeled by a measurement outcome, and in our context where one is measuring a continuous current, these POVM elements are labeled by a continuous measurement record.  

In this section we show how the Green's functions we have defined are related to a POVM on the field degrees of freedom. We consider a single mode of the field for simplicity, but the results can be generalized to more complex fields.

Using Eq.~\eqref{eq:fielddef} we obtain a recursive operator 
\begin{flalign*}
\mathcal{P}(t)=1+&\int_{-\infty}^t\bar{\mathcal{G}}(t-\tau)\\
&\hspace{0cm}\times\left(\bar{\mathcal{L}} d\Lambda_{\tau}+\bar{\mathcal{L}}^+dB_{\tau}+\bar{\mathcal{L}}^-dB^\dagger_{\tau}\right)\mathcal{P}(\tau)
\end{flalign*} 
such that our final system state $\bar{\rho}(t)$ can be expressed as
\begin{flalign*}
\bar{\rho}_{\rm MATTER}&(t)=\\
&{\rm Tr}_{\rm LIGHT}\left[\mathcal{P}(t) \bar{\rho}_{\rm MATTER}(t_0)\otimes\hat{\rho}_{\rm LIGHT}(t_0)\right]
\end{flalign*}
where $\hat{\rho}_{\rm LIGHT}(t_0)$ is our initial field state constructed as indicated above.   We can now write operator expectation values as a function of the input state 
\begin{flalign*}
\langle \hat{X}_i\rangle&(\hat{\rho}_{\rm LIGHT}(t_0),t)=\\
&{\rm Tr}_{\rm LIGHT}\left[\bar{X}_i\cdot\mathcal{P}(t) \bar{\rho}_{\rm MATTER}(t_0)\otimes\hat{\rho}_{\rm LIGHT}(t_0)\right].
\end{flalign*}

For a complete set of appropriately normalized amplification operators $\bar{X}_i$ obtained from vectorization of POVM elements $\hat{X}_i$ on the system, when an initial state is assumed, $\bar{X}_i\cdot\mathcal{P}(t) \bar{\rho}_{\rm MATTER}(t_0)$ form a coarse-grained POVM on the field.

\section{Estimation of efficiency from average dynamics}
\label{sec:eff_estimation}
The stochastic term given by Eq.~\eqref{eq:amp} unfortunately contributes both explicit time dependence and nonlinearity due to the normalizing third term of the superoperator $\mathcal{H}$, meaning that it cannot be described efficiently using the machinery presented in the main text.       
Furthermore, since detection events are defined as the output exceeding some threshold, which is a nonlinear condition on the current, one usually has to resort to stochastic trajectory simulations in order to estimate efficiencies\cite{Young:2018}. 

However, as mentioned in the main text, under commonly prevailing conditions measurement metrics can be estimated from the averaged dynamics. Under sufficiently strong amplification, the trajectories become jump-like, with the system being in either a monitored or unmonitored state.  We may thus write the probability of the system being in the monitored subsystem at time $t$ as
\begin{flalign}
P_{\rm MON}(t)&= P_{\rm MON|MON}(t,t')P_{\rm MON}(t') \nonumber \\
&+P_{\rm MON|UN,MON}(t,t')P_{\rm UN,MON}(t'),
\label{eq:Pmon_abstract}
\end{flalign}
where $P_{\rm x|y}(t,t')$ is the probability of the system being in subsystem x at $t$ given that it was in subsystem $y$ at $t'$. Here, the subsystems are labeled MON and UN,MON, which correspond to the monitored and unmonitored states.
Recall that $\hat{x}$ projects onto the subspace of states being monitored through $\hat{X}$. Therefore, $P_{\rm MON}(t)={\rm Tr}\left[\hat{x}\hat{\rho}(t)\right]=\bar{x}\dg\bar{\rho}(t)$, and hence rewriting the above,
\begin{flalign}
P_{\rm MON}(t)&= P_{\rm MON|MON}(t,t')\bar{x}\dg\bar{\rho}(t') \nonumber \\
&~~~+P_{\rm MON|UN,MON}(t,t')\left(\bv{I}-\bar{x}\right)\dg\bar{\rho}(t').
\label{eq:Pmon}
\end{flalign}

Now, we formulate a probability quantity that will serve as an approximation of detection efficiency. Consider a single detection channel for simplicity. Our proxy for efficiency is based on the time-dependent probability that a detection event is registered in the channel of interest, which we denote by $\Pi(t)$. This is the probability of registering a detection event in the interval $(t_0,t]$ given a photon in the field at time $t_0$. We write this explicitly as
\begin{flalign}
	\Pi(t) &= \int_{t_0+t_{\rm MIN}}^{t} dt' s_{\rm MON | UN,MON}(t'-t_{\rm MIN}) \times \nonumber \\
	& ~~~~~~~~~~~~~~~~~~~~~  P_{\rm MON | MON}(t',t'-t_{\rm MIN}),
	\label{eq:phit_abstract}
\end{flalign}
where $s_{\rm MON | UN,MON}(t)$ is the probability of transitioning from the unmonitored subspace to the monitored subspace at time $t$. The integrand in this expression represents the joint probability of population transferring from the unmonitored subspace (which includes the field degrees of freedom) to the monitored subspace at time $t'$, and that this population stays in the monitored subspace for time $t_{\rm MIN}$, which we take as the minimum time that the monitored subspace has to be populated before a detection is registered. (A more experimentally motivated condition for registering a detection is that the measured current exceed some threshold value $I_{\rm HIT}$; however, since it is difficult to develop an expression based on this condition we proceed with the temporal condition. The approximate relationship between the two is that $t_{\rm MON} \approx I_{\rm HIT}t_m/\chi$.)

In order to proceed we require an expression for $s_{\rm MON | UN,MON}(t)$. To obtain this, we return to 
Eq.~(\ref{eq:Pmon_abstract}) and consider the incremental quantity
\begin{flalign*}
	\Delta P_{\rm MON}(t) &\equiv P_{\rm MON}(t) - P_{\rm MON}(t-\Delta t)  \\
	&= \Big[ P_{\rm MON|MON}(t,t- \Delta t) - 1\Big]P_{\rm MON} (t- \Delta t) \\
	& ~~~+ P_{\rm MON|UN,MON}(t,t-\Delta t)P_{\rm UN,MON}(t-\Delta t).
\end{flalign*}
Now, the first quantity in this expression corresponds to the rate at which population is staying in the monitored subspace while the second quantity corresponds to the rate at which population is entering the monitored subspace from the unmonitored subspace at time $t$. This is precisely the quantity $s_{\rm MON | UN,MON}(t) \Delta t$, and given this expression for it, we can compute it explicitly in terms of the density matrix and dynamics:
\begin{flalign*}
&s_{\rm MON | UN,MON}(t)\nonumber \\
&=\frac{d}{dt} \Bigg(P_{\rm MON|UN,MON}(t,t')P_{\rm UN,MON} (t-dt) \Bigg)\vert_{t'=t} \nonumber \\
&=\frac{d}{dt} \Bigg(P_{\rm MON|UN,MON}(t,t')\left(\bv{I}-\bar{x}\right)\dg\bar{\rho}(t') \Bigg)\vert_{t'=t} \nonumber \\
&=\frac{d}{dt} \Bigg( \bar{x}\dg \bar{\rho}(t) - P_{\rm MON|MON}(t,t') \bar{x}\dg\bar{\rho}(t')\Bigg) \vert_{t'=t} \nonumber \\
&=\bar{x}\dg\left(\frac{d\bar{\rho} (t)}{dt}-\frac{dP_{\rm MON|MON}(t,t')}{dt}\vert_{t'=t}\bar{\rho}(t)\right),
\end{flalign*}
where in the third line we have used the identity in Eq. (\ref{eq:Pmon}). 
$P_{\rm MON|MON}(t,t')$ is typically a function of the internal dynamics and can then be readily determined from the system $\bar{\mathcal{G}}$. When $\bar{\mathcal{A}}$ is time-independent, it can further be written as a function of $\tau = t-t'$ as
\begin{flalign*}
P_{\rm MON|MON}(t,t')=\bar{x}^{\dagger}\bar{\mathcal{G}}(\tau)\bar{x}.
\end{flalign*}
Hence,
\begin{flalign*}
\frac{dP_{\rm MON|MON}(t,t')}{dt}\vert_{t=t'}&=\bar{x}^{\dagger}\frac{d\bar{\mathcal{G}}(t)}{dt}\vert_{t=0}\bar{x}\\
&=\bar{x}^{\dagger}\bar{\mathcal{A}}\bar{x}.
\end{flalign*}
Finally, using this expression for $s_{\rm MON | UN,MON}(t)$ in Eq. (\ref{eq:phit_abstract}), results in
\begin{flalign*}
	\Pi(t) &= \int_{t_0+t_{\rm MON}}^{t} dt' ~s_{\rm MON | UN,MON}(t'-t_{\rm MON}) \times \nonumber \\
	& ~~~~~~~~~~~~~~~~~~~~~  P_{\rm MON | MON}(t',t'-t_{\rm MON}) \\
	&= \int_{t_0}^{t-t_{\rm MON}} d\tau ~s_{\rm MON | UN,MON}(\tau) \times \nonumber \\
	& ~~~~~~~~~~~~~~~~~~~~~  P_{\rm MON | MON}(\tau+t_{\rm MON},\tau) \\
	&= \int_{t_0}^{t-t_{\rm MON}} d\tau ~P_{\rm MON|MON}(t_{\rm MON}) \times \nonumber \\
	& ~~~~~~~~~~~~~~~~~~~\bar{x}\dg\left(\frac{d\bar{\rho} (\tau)}{d\tau}-\bar{x}^{\dagger}\bar{\mathcal{A}}\bar{x}\bar{\rho}(\tau)\right) \\
	&= \bar{x}\dg \bar{\mathcal{G}}(t_{\rm MON})\bar{x} \Big(\bar{x}\dg\bar{\rho} (t-t_{\rm MON}) - \Big.\\
	&~~~~~~~~~~~~~~~~~~~~~~ \Big. (\bar{x}^{\dagger}\bar{\mathcal{A}}\bar{x}) \bar{x}\dg\int_{t_0}^{t-t_{\rm MON}} d\tau~ \bar{\rho}(\tau)\Big),
\end{flalign*}
where we have assumed that $\bar{x}\dg \bar{\rho}(t_0)=0$.

If the monitored subsystem is stable and population loss from it can be neglected, $P_{\rm MON|MON}(t,t')=1$, and therefore $\bar{x}^{\dagger}\bar{\mathcal{G}}\bar{x}=1$ and $\bar{x}^{\dagger} A \bar{x}=0$. In this case, the above reduces to 
\begin{flalign}
\Pi(t)&=\bar{x} \dg \bar{\rho}(t-t_{\rm MON})\label{eq:Pishelve}.
\end{flalign}
These expressions are the probability that a detection event is registered in the measurement channel.
In general, there may be multiple measurement channels, each associated with an operator $\hat{X}_i$, projection subspace $\hat{x}_i$, and $\Pi_i(t)$.  The probability that a given number of photons $N$ are detected will depend in some fashion on the combined behavior of all the channels, so that $P_N$ is some function $p_N$ of all $\Pi_i$.  

As an example, we refer to the system discussed in Section \ref{sec:other}. Since each element has an associated measurement channel, we must consider $\Pi_i, i\in 1..n$.  Since the absorption of a photon corresponds to one measurement channel registering a hit, we take the registering of a hit on $N$ channels to correspond to detection of an $N$ photon wavepacket. Thus, for a single photon, $p_1(\Pi_i)=\sum_i \Pi_i$, and for, \eg three photons, $p_3(\Pi_i)=\sum_{i> j> k} \Pi_i|\Pi_j|\Pi_k$, where $\Pi_i|\Pi_j$ is the probability of a detection event on the $i$th channel given one on the $j$th channel.  Since this system satisfies the conditions of Eq.~\eqref{eq:Pishelve}, this can be expressed as 
\begin{flalign*}
\Pi_i|\Pi_j|\Pi_k...=\left(\bar{x}_i\times\bar{x}_j\times\bar{x}_k...\right)\dg \bar{\rho}(t-t_{\rm MON})
\end{flalign*} 
where $\times$ represents an element-wise product.

\section{Degenerate Multi Element}
\label{sec:degenerate}
For a system of $n$ elements, we can represent the total system density matrix using the notation
\begin{flalign*}
\hat{\rho}_{n:m_il_im_jl_jm_kl_k...}=\hat{\rho}^i_{m_il_i}\otimes\hat{\rho}^j_{m_jl_j}\otimes\hat{\rho}^k_{m_kl_k}\otimes...
\end{flalign*}
where $m_i$ and $l_i$ are one of $0_i,1_i,C_i$, corresponding to the states of subsystem $i$. For example, $\hat{\rho}_{3:0_10_11_20_20_31_3}$  represents a system with $3$ subsystems and coherences between the 0 and 1 states in subsystems 2 and 3. 
Since the subsystems are all degenerate, for brevity only the subsystems not in the ground state will be noted in the subscript and we use, for example, $\hat{\rho}_{n:1001}$ to represent all of the $n(n-1)/2$ configurations with this description.
The operators on the vectorized density matrix are then
\begin{widetext}
\begin{flalign*}
\bar{\rho}&=\left[\begin{array}{c}
\hat{\rho}_{00}\\
\hat{\rho}_{n:11}\\
\hat{\rho}_{n:1001}\\
\hat{\rho}_{n:CC}\\
\vdots\\
\hat{\rho}_{n:10}\\
\hat{\rho}_{n:01}\\
\vdots\\
\end{array}\right]
\end{flalign*}
so that
\begin{flalign*}
\bar{\mathcal{A}}=\left[\begin{array}{c c c c c c c c c}
0 & n\gamma^2            & 0 & n(n-1)\gamma^2 &        & 0 & 0 &\\ 
0 & -(\gamma^2+\Gamma^2) & 0 & -(n-1)\gamma^2 &        & 0 & 0 & \\
0 & \Gamma^2             & 0 & 0              & \hdots & 0 & 0 &\hdots\\
0 & -\gamma^2            & 0 & -(n-1)\gamma^2 &        & 0 & 0& \\
 &  & \vdots& & \ddots & & \\
0 & 0 & 0 & 0 &  & i\omega-\frac{n\gamma^2+\Gamma^2}{2} & 0 &  \\
0 & 0 & 0 & 0 &  & 0 & -i\omega-\frac{n\gamma^2+\Gamma^2}{2} &  \\
 &  & \vdots& & & & &\ddots 
\end{array}\right],\\
\bar{\mathcal{L}}^+=\left[\begin{array}{c c c c c c c c c}
0 & 0           & 0 &0  &        & -n\gamma & 0 & \\ 
0 & 0           & 0 &0  &        & \gamma & 0 & \\ 
0 & 0           & 0 &0  &  \hdots      & (n-1)\gamma & 0 & \hdots\\ 
0 & 0           & 0 &0  &        & 0 & 0 & \\ 
&  & \vdots& & \ddots & \\
0 & 0 & 0 & 0 &  & 0 & 0 &   \\
\gamma & -\gamma & -(n-1)\gamma & 0 &  & 0 & 0 &   \\
&  & \vdots& & & & &\ddots & 
\end{array}\right]
\end{flalign*}

\end{widetext}

\section{Dispersive Band}
Much of the machinery developed in other contexts for manipulating Green's functions may be brought to bear as well.  In some cases it is convenient to separate $\bar{\mathcal{A}}$ into an $\bar{\mathcal{A}}^0$ and $\bar{\mathcal{A}}^1$, where the former can be easily exponentiated (e.g., it is diagonal) so that    
\begin{flalign}
\bar{\mathcal{G}}(t)&=\expn{\bar{\mathcal{A}}t}=\expn{\left(\bar{\mathcal{A}}^0t+\bar{\mathcal{A}}^1t\right) }=\lim_{dt\rightarrow 0}\left[\expn{\bar{\mathcal{A}}^0dt}+\bar{\mathcal{A}}^1 dt\right]^{t/dt}\nonumber\\
&=\hat{g}(t)+\int_{0}^t d\tau\hat{g} (t-\tau)\bar{\mathcal{A}}^1\hat{g}(\tau)\nonumber\\
&+\int_{-\infty}^t d\tau\hat{g} (t-\tau)\bar{\mathcal{A}}^1\int_{0}^\tau d\tau'\hat{g} (\tau-\tau')\bar{\mathcal{A}}^1\hat{g}(\tau')+ ...\nonumber\\
&=\hat{g}(t)+\int_{0}^t d\tau\hat{g} (t-\tau)\bar{\mathcal{A}}^1\bar{\mathcal{G}}(\tau)\label{eq:gfseries}.
\end{flalign}

An example of this is provided by a system comprising a ground state labeled 0 and a dispersive band of $n$ excited states labeled with the indices $1..n$.  All excitations are characterized by the same light-matter coupling $\gamma$. For clarity we represent elements of $\bar{\rho}$ with the notation ${\bar{\rho}}_{(ij)}=\hat{\rho}_{ij}$ and the element of a matrix $\bar{\mathcal{O}}$ coupling density matrix elements $\hat{\rho_{ij}}$ and $\hat{\rho}_{rs}$ with the notation $\bar{\mathcal{O}}_{(ij)(rs)}$. 

Then $\bar{\mathcal{A}}^0$ and $\bar{\mathcal{A}}^1$ are, giving only nonzero elements, 
\begin{flalign*}
\begin{array}{l r}
\bar{\mathcal{A}}^0_{(ij)(ij)}=i(\omega_i-\omega_j) &i,j\in 0..n\\
&\\
\bar{\mathcal{A}}^1_{(00)(ij)}=\gamma^2 &\\
\bar{\mathcal{A}}^1_{(0i)(0j)}=\bar{\mathcal{A}}^1_{(i0)(j0)}=-\gamma^2/2 &\hspace{2cm} i,j,r,s\in 1..n\\
\bar{\mathcal{A}}^1_{(ij)(rs)}=-\frac{\gamma^2}{2}\left(\delta_{ir}+\delta_{js}\right) &
\end{array}
\end{flalign*}

The total population of the band is given by 
\begin{flalign*}
u(t)&=\sum_{i}\bar{\varrho}^{11}_{(ii)}(t)\\
&=\sum_{ijrs} 2\int_{-\infty}^t d\tau \gamma\bar{\mathcal{G}}_{(ii)(rj)}(t-\tau)E(\tau)\\
&\times\int_{-\infty}^t d\tau \bar{\mathcal{G}}_{(r0)(s0)}(t-\tau)E(\tau)P0_{(00)}\\
\end{flalign*}
Expanding $\sum_{s}\bar{\mathcal{G}}_{(r0)(s0)}$ according to  Eq.~\eqref{eq:gfseries}
\begin{flalign*}
\sum_{s}\bar{\mathcal{G}}_{(r0)(s0)}&(t)\\
&=\expn{-i\omega_r t}-\sum_{r}\frac{\gamma^2}{2}\int_{0}^td\tau \expn{-i\omega_r(t-\tau)}\expn{-i\omega_s\tau}\\
&+\sum_{qs}\frac{\gamma^4}{4}\int_{0}^td\tau \expn{-i\omega_r(t-\tau)}\\
&\times\int_0^\tau d\tau'\expn{-i\omega_q(\tau-\tau')}\expn{-i\omega_s\tau'}+...
\end{flalign*}
If we take the continuum limit $n\rightarrow\infty$ and assume a Lorentzian distribution of states $D(\omega)=\frac{n}{\pi}\frac{\zeta^2/2}{(\zeta^2/2)^2+(\omega-\omega_E)^2}$ around a center frequency $\omega_E$ then $\sum_{r}\expn{i\omega_rt}=\int\sigma(\omega_r)d\omega_r\expn{i\omega_rt}=n\expn{-(\zeta^2/2)t}$ and the series can be evaluated analytically as
\begin{flalign*}
\sum_{s}\bar{\mathcal{G}}_{(r0)(s0)}&(t)\\
&=\expn{-i\omega_r t}-\sum_{s}\frac{\gamma^2}{2}\int_{0}^td\tau \expn{-i\omega_r(t-\tau)}n\expn{-(\zeta^2/2)\tau}\\
&+\sum_{qs}\frac{\gamma^4}{4}\int_{0}^td\tau \expn{-i\omega_r(t-\tau)}\int_0^\tau d\tau'n\expn{-(\zeta^2/2)\tau}+...\\
&=\expn{-i\omega_r t}-\frac{n\gamma^2}{2}\int_{0}^td\tau \expn{-i\omega_r(t-\tau)}\\
&\times\expn{-\zeta^2/2\tau}\left(1-\frac{n\gamma^2}{2}+\frac{n^2\gamma^4}{4}...\right)\\
&=\expn{-i\omega_r t}-\frac{n\gamma^2}{2}\int_{0}^td\tau \expn{-i\omega_r(t-\tau)}\expn{-\frac{\zeta^2+n\gamma^2}{2}\tau}.
\end{flalign*}
Following a similar procedure for $\sum_r\bar{\mathcal{G}}_{(pq)(r0)}(t)$, we can ultimately write
\begin{flalign*}
u(t)&=\frac{ 2\gamma^2}{n\gamma^2+\zeta^2}\int^t d\tau \left(n\gamma^2\expn{-(n\gamma^2+\zeta^2)(t-\tau)}+\zeta^2\right)\\
&\times E(\tau)\int^\tau d\tau'  \expn{-\frac{n\gamma^2+\zeta^2}{2}(\tau-\tau')} E(\tau').
\end{flalign*}

\clearpage

\bibliography{stoch_draft}

\begin{thebibliography}{29}%
\makeatletter
\providecommand \@ifxundefined [1]{%
 \@ifx{#1\undefined}
}%
\providecommand \@ifnum [1]{%
 \ifnum #1\expandafter \@firstoftwo
 \else \expandafter \@secondoftwo
 \fi
}%
\providecommand \@ifx [1]{%
 \ifx #1\expandafter \@firstoftwo
 \else \expandafter \@secondoftwo
 \fi
}%
\providecommand \natexlab [1]{#1}%
\providecommand \enquote  [1]{``#1''}%
\providecommand \bibnamefont  [1]{#1}%
\providecommand \bibfnamefont [1]{#1}%
\providecommand \citenamefont [1]{#1}%
\providecommand \href@noop [0]{\@secondoftwo}%
\providecommand \href [0]{\begingroup \@sanitize@url \@href}%
\providecommand \@href[1]{\@@startlink{#1}\@@href}%
\providecommand \@@href[1]{\endgroup#1\@@endlink}%
\providecommand \@sanitize@url [0]{\catcode `\\12\catcode `\$12\catcode
  `\&12\catcode `\#12\catcode `\^12\catcode `\_12\catcode `\%12\relax}%
\providecommand \@@startlink[1]{}%
\providecommand \@@endlink[0]{}%
\providecommand \url  [0]{\begingroup\@sanitize@url \@url }%
\providecommand \@url [1]{\endgroup\@href {#1}{\urlprefix }}%
\providecommand \urlprefix  [0]{URL }%
\providecommand \Eprint [0]{\href }%
\providecommand \doibase [0]{http://dx.doi.org/}%
\providecommand \selectlanguage [0]{\@gobble}%
\providecommand \bibinfo  [0]{\@secondoftwo}%
\providecommand \bibfield  [0]{\@secondoftwo}%
\providecommand \translation [1]{[#1]}%
\providecommand \BibitemOpen [0]{}%
\providecommand \bibitemStop [0]{}%
\providecommand \bibitemNoStop [0]{.\EOS\space}%
\providecommand \EOS [0]{\spacefactor3000\relax}%
\providecommand \BibitemShut  [1]{\csname bibitem#1\endcsname}%
\let\auto@bib@innerbib\@empty
%</preamble>
\bibitem [{\citenamefont {Bienfang}\ \emph {et~al.}(2004)\citenamefont
  {Bienfang}, \citenamefont {Gross}, \citenamefont {Mink}, \citenamefont
  {Hershman}, \citenamefont {Nakassis}, \citenamefont {Tang}, \citenamefont
  {Lu}, \citenamefont {Su}, \citenamefont {Clark}, \citenamefont {Williams},
  \citenamefont {Hagley},\ and\ \citenamefont {Wen}}]{Bienfang:2004ij}%
  \BibitemOpen
  \bibfield  {author} {\bibinfo {author} {\bibfnamefont {J.~C.}\ \bibnamefont
  {Bienfang}}, \bibinfo {author} {\bibfnamefont {A.~J.}\ \bibnamefont {Gross}},
  \bibinfo {author} {\bibfnamefont {A.}~\bibnamefont {Mink}}, \bibinfo {author}
  {\bibfnamefont {B.~J.}\ \bibnamefont {Hershman}}, \bibinfo {author}
  {\bibfnamefont {A.}~\bibnamefont {Nakassis}}, \bibinfo {author}
  {\bibfnamefont {X.}~\bibnamefont {Tang}}, \bibinfo {author} {\bibfnamefont
  {R.}~\bibnamefont {Lu}}, \bibinfo {author} {\bibfnamefont {D.~H.}\
  \bibnamefont {Su}}, \bibinfo {author} {\bibfnamefont {C.~W.}\ \bibnamefont
  {Clark}}, \bibinfo {author} {\bibfnamefont {C.~J.}\ \bibnamefont {Williams}},
  \bibinfo {author} {\bibfnamefont {E.~W.}\ \bibnamefont {Hagley}}, \ and\
  \bibinfo {author} {\bibfnamefont {J.}~\bibnamefont {Wen}},\ }\href@noop {}
  {\bibfield  {journal} {\bibinfo  {journal} {Opt. Express}\ }\textbf {\bibinfo
  {volume} {12}},\ \bibinfo {pages} {2011} (\bibinfo {year}
  {2004})}\BibitemShut {NoStop}%
\bibitem [{\citenamefont {Woodson}\ \emph {et~al.}(2016)\citenamefont
  {Woodson}, \citenamefont {Ren}, \citenamefont {Maddox}, \citenamefont {Chen},
  \citenamefont {Bank},\ and\ \citenamefont {Campbell}}]{Woodson:2016cx}%
  \BibitemOpen
  \bibfield  {author} {\bibinfo {author} {\bibfnamefont {M.~E.}\ \bibnamefont
  {Woodson}}, \bibinfo {author} {\bibfnamefont {M.}~\bibnamefont {Ren}},
  \bibinfo {author} {\bibfnamefont {S.~J.}\ \bibnamefont {Maddox}}, \bibinfo
  {author} {\bibfnamefont {Y.}~\bibnamefont {Chen}}, \bibinfo {author}
  {\bibfnamefont {S.~R.}\ \bibnamefont {Bank}}, \ and\ \bibinfo {author}
  {\bibfnamefont {J.~C.}\ \bibnamefont {Campbell}},\ }\href@noop {} {\bibfield
  {journal} {\bibinfo  {journal} {Appl. Phys. Lett.}\ }\textbf {\bibinfo
  {volume} {108}},\ \bibinfo {pages} {081102} (\bibinfo {year}
  {2016})}\BibitemShut {NoStop}%
\bibitem [{\citenamefont {Pernice}\ \emph {et~al.}(2012)\citenamefont
  {Pernice}, \citenamefont {Schuck}, \citenamefont {Minaeva}, \citenamefont
  {Li}, \citenamefont {Goltsman}, \citenamefont {Sergienko},\ and\
  \citenamefont {Tang}}]{Pernice:2012bc}%
  \BibitemOpen
  \bibfield  {author} {\bibinfo {author} {\bibfnamefont {W.~H.~P.}\
  \bibnamefont {Pernice}}, \bibinfo {author} {\bibfnamefont {C.}~\bibnamefont
  {Schuck}}, \bibinfo {author} {\bibfnamefont {O.}~\bibnamefont {Minaeva}},
  \bibinfo {author} {\bibfnamefont {M.}~\bibnamefont {Li}}, \bibinfo {author}
  {\bibfnamefont {G.~N.}\ \bibnamefont {Goltsman}}, \bibinfo {author}
  {\bibfnamefont {A.~V.}\ \bibnamefont {Sergienko}}, \ and\ \bibinfo {author}
  {\bibfnamefont {H.~X.}\ \bibnamefont {Tang}},\ }\href@noop {} {\bibfield
  {journal} {\bibinfo  {journal} {Nature}\ }\textbf {\bibinfo {volume} {3}},\
  \bibinfo {pages} {1325} (\bibinfo {year} {2012})}\BibitemShut {NoStop}%
\bibitem [{\citenamefont {Marsili}\ \emph {et~al.}(2012)\citenamefont
  {Marsili}, \citenamefont {Bellei}, \citenamefont {Najafi}, \citenamefont
  {Dane}, \citenamefont {Dauler}, \citenamefont {Molnar},\ and\ \citenamefont
  {Berggren}}]{Marsili:2012ib}%
  \BibitemOpen
  \bibfield  {author} {\bibinfo {author} {\bibfnamefont {F.}~\bibnamefont
  {Marsili}}, \bibinfo {author} {\bibfnamefont {F.}~\bibnamefont {Bellei}},
  \bibinfo {author} {\bibfnamefont {F.}~\bibnamefont {Najafi}}, \bibinfo
  {author} {\bibfnamefont {A.~E.}\ \bibnamefont {Dane}}, \bibinfo {author}
  {\bibfnamefont {E.~A.}\ \bibnamefont {Dauler}}, \bibinfo {author}
  {\bibfnamefont {R.~J.}\ \bibnamefont {Molnar}}, \ and\ \bibinfo {author}
  {\bibfnamefont {K.~K.}\ \bibnamefont {Berggren}},\ }\href@noop {} {\bibfield
  {journal} {\bibinfo  {journal} {Nano. Lett.}\ }\textbf {\bibinfo {volume}
  {12}},\ \bibinfo {pages} {4799} (\bibinfo {year} {2012})}\BibitemShut
  {NoStop}%
\bibitem [{\citenamefont {Marsili}\ \emph {et~al.}(2013)\citenamefont
  {Marsili}, \citenamefont {Verma}, \citenamefont {Stern}, \citenamefont
  {Harrington}, \citenamefont {Lita}, \citenamefont {Gerrits}, \citenamefont
  {Vayshenker}, \citenamefont {Baek}, \citenamefont {Shaw}, \citenamefont
  {Mirin},\ and\ \citenamefont {Nam}}]{Marsili:2013th}%
  \BibitemOpen
  \bibfield  {author} {\bibinfo {author} {\bibfnamefont {F.}~\bibnamefont
  {Marsili}}, \bibinfo {author} {\bibfnamefont {V.~B.}\ \bibnamefont {Verma}},
  \bibinfo {author} {\bibfnamefont {J.~A.}\ \bibnamefont {Stern}}, \bibinfo
  {author} {\bibfnamefont {S.}~\bibnamefont {Harrington}}, \bibinfo {author}
  {\bibfnamefont {A.~E.}\ \bibnamefont {Lita}}, \bibinfo {author}
  {\bibfnamefont {T.}~\bibnamefont {Gerrits}}, \bibinfo {author} {\bibfnamefont
  {I.}~\bibnamefont {Vayshenker}}, \bibinfo {author} {\bibfnamefont
  {B.}~\bibnamefont {Baek}}, \bibinfo {author} {\bibfnamefont {M.~D.}\
  \bibnamefont {Shaw}}, \bibinfo {author} {\bibfnamefont {R.~P.}\ \bibnamefont
  {Mirin}}, \ and\ \bibinfo {author} {\bibfnamefont {S.~W.}\ \bibnamefont
  {Nam}},\ }\href@noop {} {\bibfield  {journal} {\bibinfo  {journal} {Nat.
  Photonics}\ }\textbf {\bibinfo {volume} {7}},\ \bibinfo {pages} {210}
  (\bibinfo {year} {2013})}\BibitemShut {NoStop}%
\bibitem [{\citenamefont {Eisaman}\ \emph {et~al.}(2011)\citenamefont
  {Eisaman}, \citenamefont {Fan}, \citenamefont {Migdall},\ and\ \citenamefont
  {Polyakov}}]{Eisaman:2011cc}%
  \BibitemOpen
  \bibfield  {author} {\bibinfo {author} {\bibfnamefont {M.~D.}\ \bibnamefont
  {Eisaman}}, \bibinfo {author} {\bibfnamefont {J.}~\bibnamefont {Fan}},
  \bibinfo {author} {\bibfnamefont {A.}~\bibnamefont {Migdall}}, \ and\
  \bibinfo {author} {\bibfnamefont {S.~V.}\ \bibnamefont {Polyakov}},\
  }\href@noop {} {\bibfield  {journal} {\bibinfo  {journal} {Review of
  Scientific Instruments}\ }\textbf {\bibinfo {volume} {82}},\ \bibinfo {pages}
  {071101} (\bibinfo {year} {2011})}\BibitemShut {NoStop}%
\bibitem [{\citenamefont {Hadfield}(2009)}]{Hadfield:2009}%
  \BibitemOpen
  \bibfield  {author} {\bibinfo {author} {\bibfnamefont {R.~H.}\ \bibnamefont
  {Hadfield}},\ }\href@noop {} {\bibfield  {journal} {\bibinfo  {journal} {Nat.
  Photonics}\ }\textbf {\bibinfo {volume} {3}},\ \bibinfo {pages} {696}
  (\bibinfo {year} {2009})}\BibitemShut {NoStop}%
\bibitem [{\citenamefont {Cassemiro}\ \emph {et~al.}(2010)\citenamefont
  {Cassemiro}, \citenamefont {Laiho},\ and\ \citenamefont
  {Silberhorn}}]{Cassemiro:2010}%
  \BibitemOpen
  \bibfield  {author} {\bibinfo {author} {\bibfnamefont {K.~N.}\ \bibnamefont
  {Cassemiro}}, \bibinfo {author} {\bibfnamefont {K.}~\bibnamefont {Laiho}}, \
  and\ \bibinfo {author} {\bibfnamefont {C.}~\bibnamefont {Silberhorn}},\
  }\href {http://stacks.iop.org/1367-2630/12/i=11/a=113052} {\bibfield
  {journal} {\bibinfo  {journal} {New J. Phys.}\ }\textbf {\bibinfo {volume}
  {12}},\ \bibinfo {pages} {113052} (\bibinfo {year} {2010})}\BibitemShut
  {NoStop}%
\bibitem [{\citenamefont {Piacentini}\ \emph {et~al.}(2017)\citenamefont
  {Piacentini}, \citenamefont {Avella}, \citenamefont {Rebufello},
  \citenamefont {Lussana}, \citenamefont {Villa}, \citenamefont {Tosi},
  \citenamefont {Gramegna}, \citenamefont {Brida}, \citenamefont {Cohen},
  \citenamefont {Vaidman}, \citenamefont {Degiovanni},\ and\ \citenamefont
  {Genovese}}]{Genovese:2017}%
  \BibitemOpen
  \bibfield  {author} {\bibinfo {author} {\bibfnamefont {F.}~\bibnamefont
  {Piacentini}}, \bibinfo {author} {\bibfnamefont {A.}~\bibnamefont {Avella}},
  \bibinfo {author} {\bibfnamefont {E.}~\bibnamefont {Rebufello}}, \bibinfo
  {author} {\bibfnamefont {R.}~\bibnamefont {Lussana}}, \bibinfo {author}
  {\bibfnamefont {F.}~\bibnamefont {Villa}}, \bibinfo {author} {\bibfnamefont
  {A.}~\bibnamefont {Tosi}}, \bibinfo {author} {\bibfnamefont {M.}~\bibnamefont
  {Gramegna}}, \bibinfo {author} {\bibfnamefont {G.}~\bibnamefont {Brida}},
  \bibinfo {author} {\bibfnamefont {E.}~\bibnamefont {Cohen}}, \bibinfo
  {author} {\bibfnamefont {L.}~\bibnamefont {Vaidman}}, \bibinfo {author}
  {\bibfnamefont {I.~P.}\ \bibnamefont {Degiovanni}}, \ and\ \bibinfo {author}
  {\bibfnamefont {M.}~\bibnamefont {Genovese}},\ }\href@noop {} {\bibfield
  {journal} {\bibinfo  {journal} {Nat. Photonics}\ }\textbf {\bibinfo {volume}
  {13}},\ \bibinfo {pages} {1191} (\bibinfo {year} {2017})}\BibitemShut
  {NoStop}%
\bibitem [{\citenamefont {\ifmmode \check{R}\else
  \v{R}\fi{}eh\'a\ifmmode~\check{c}\else \v{c}\fi{}ek}\ \emph
  {et~al.}(2003)\citenamefont {\ifmmode \check{R}\else
  \v{R}\fi{}eh\'a\ifmmode~\check{c}\else \v{c}\fi{}ek}, \citenamefont {Hradil},
  \citenamefont {Haderka}, \citenamefont {Pe\ifmmode~\check{r}\else
  \v{r}\fi{}ina},\ and\ \citenamefont {Hamar}}]{Perina:2003}%
  \BibitemOpen
  \bibfield  {author} {\bibinfo {author} {\bibfnamefont {J.}~\bibnamefont
  {\ifmmode \check{R}\else \v{R}\fi{}eh\'a\ifmmode~\check{c}\else
  \v{c}\fi{}ek}}, \bibinfo {author} {\bibfnamefont {Z.}~\bibnamefont {Hradil}},
  \bibinfo {author} {\bibfnamefont {O.}~\bibnamefont {Haderka}}, \bibinfo
  {author} {\bibfnamefont {J.}~\bibnamefont {Pe\ifmmode~\check{r}\else
  \v{r}\fi{}ina}}, \ and\ \bibinfo {author} {\bibfnamefont {M.}~\bibnamefont
  {Hamar}},\ }\href {\doibase 10.1103/PhysRevA.67.061801} {\bibfield  {journal}
  {\bibinfo  {journal} {Phys. Rev. A}\ }\textbf {\bibinfo {volume} {67}},\
  \bibinfo {pages} {061801} (\bibinfo {year} {2003})}\BibitemShut {NoStop}%
\bibitem [{\citenamefont {Achilles}\ \emph {et~al.}(2003)\citenamefont
  {Achilles}, \citenamefont {Silberhorn}, \citenamefont {\'{S}liwa},
  \citenamefont {Banaszek},\ and\ \citenamefont {Walmsley}}]{Achilles:03}%
  \BibitemOpen
  \bibfield  {author} {\bibinfo {author} {\bibfnamefont {D.}~\bibnamefont
  {Achilles}}, \bibinfo {author} {\bibfnamefont {C.}~\bibnamefont
  {Silberhorn}}, \bibinfo {author} {\bibfnamefont {C.}~\bibnamefont
  {\'{S}liwa}}, \bibinfo {author} {\bibfnamefont {K.}~\bibnamefont {Banaszek}},
  \ and\ \bibinfo {author} {\bibfnamefont {I.~A.}\ \bibnamefont {Walmsley}},\
  }\href@noop {} {\bibfield  {journal} {\bibinfo  {journal} {Opt. Lett.}\
  }\textbf {\bibinfo {volume} {28}},\ \bibinfo {pages} {2387} (\bibinfo {year}
  {2003})}\BibitemShut {NoStop}%
\bibitem [{\citenamefont {Sperling}\ \emph {et~al.}(2016)\citenamefont
  {Sperling}, \citenamefont {Bartley}, \citenamefont {Donati}, \citenamefont
  {Barbieri}, \citenamefont {Jin}, \citenamefont {Datta}, \citenamefont
  {Vogel},\ and\ \citenamefont {Walmsley}}]{Wamsley:2016}%
  \BibitemOpen
  \bibfield  {author} {\bibinfo {author} {\bibfnamefont {J.}~\bibnamefont
  {Sperling}}, \bibinfo {author} {\bibfnamefont {T.~J.}\ \bibnamefont
  {Bartley}}, \bibinfo {author} {\bibfnamefont {G.}~\bibnamefont {Donati}},
  \bibinfo {author} {\bibfnamefont {M.}~\bibnamefont {Barbieri}}, \bibinfo
  {author} {\bibfnamefont {X.-M.}\ \bibnamefont {Jin}}, \bibinfo {author}
  {\bibfnamefont {A.}~\bibnamefont {Datta}}, \bibinfo {author} {\bibfnamefont
  {W.}~\bibnamefont {Vogel}}, \ and\ \bibinfo {author} {\bibfnamefont {I.~A.}\
  \bibnamefont {Walmsley}},\ }\href {\doibase 10.1103/PhysRevLett.117.083601}
  {\bibfield  {journal} {\bibinfo  {journal} {Phys. Rev. Lett.}\ }\textbf
  {\bibinfo {volume} {117}},\ \bibinfo {pages} {083601} (\bibinfo {year}
  {2016})}\BibitemShut {NoStop}%
\bibitem [{\citenamefont {Glauber}(1963)}]{glauber_1963}%
  \BibitemOpen
  \bibfield  {author} {\bibinfo {author} {\bibfnamefont {R.~J.}\ \bibnamefont
  {Glauber}},\ }\href@noop {} {\bibfield  {journal} {\bibinfo  {journal} {Phys.
  Rev.}\ }\textbf {\bibinfo {volume} {130}},\ \bibinfo {pages} {2529} (\bibinfo
  {year} {1963})}\BibitemShut {NoStop}%
\bibitem [{\citenamefont {Mandel}\ \emph {et~al.}(1964)\citenamefont {Mandel},
  \citenamefont {Sudarshan},\ and\ \citenamefont {Wolf}}]{mandel_1964}%
  \BibitemOpen
  \bibfield  {author} {\bibinfo {author} {\bibfnamefont {L.}~\bibnamefont
  {Mandel}}, \bibinfo {author} {\bibfnamefont {E.~C.~G.}\ \bibnamefont
  {Sudarshan}}, \ and\ \bibinfo {author} {\bibfnamefont {E.}~\bibnamefont
  {Wolf}},\ }\href@noop {} {\bibfield  {journal} {\bibinfo  {journal} {Proc.
  Phys. Soc.}\ }\textbf {\bibinfo {volume} {84}},\ \bibinfo {pages} {435}
  (\bibinfo {year} {1964})}\BibitemShut {NoStop}%
\bibitem [{\citenamefont {Kelley}\ and\ \citenamefont
  {Kleiner}(1964)}]{kelley_1964}%
  \BibitemOpen
  \bibfield  {author} {\bibinfo {author} {\bibfnamefont {P.~L.}\ \bibnamefont
  {Kelley}}\ and\ \bibinfo {author} {\bibfnamefont {W.~H.}\ \bibnamefont
  {Kleiner}},\ }\href@noop {} {\bibfield  {journal} {\bibinfo  {journal} {Phys.
  Rev.}\ }\textbf {\bibinfo {volume} {136}},\ \bibinfo {pages} {A316} (\bibinfo
  {year} {1964})}\BibitemShut {NoStop}%
\bibitem [{\citenamefont {Scully}\ and\ \citenamefont
  {LAMB}(1969)}]{scully_quantum_1969}%
  \BibitemOpen
  \bibfield  {author} {\bibinfo {author} {\bibfnamefont {M.~O.}\ \bibnamefont
  {Scully}}\ and\ \bibinfo {author} {\bibfnamefont {W.~E.}\ \bibnamefont
  {LAMB}},\ }\href {\doibase 10.1103/PhysRev.179.368} {\bibfield  {journal}
  {\bibinfo  {journal} {Phys. Rev.}\ }\textbf {\bibinfo {volume} {179}},\
  \bibinfo {pages} {368} (\bibinfo {year} {1969})}\BibitemShut {NoStop}%
\bibitem [{\citenamefont {Srinivas}\ and\ \citenamefont
  {Davies}(1981)}]{srinivas_photon_1981}%
  \BibitemOpen
  \bibfield  {author} {\bibinfo {author} {\bibfnamefont {M.~D.}\ \bibnamefont
  {Srinivas}}\ and\ \bibinfo {author} {\bibfnamefont {E.~B.}\ \bibnamefont
  {Davies}},\ }\href {\doibase 10.1080/713820643} {\bibfield  {journal}
  {\bibinfo  {journal} {Opt. Acta}\ }\textbf {\bibinfo {volume} {28}},\
  \bibinfo {pages} {981} (\bibinfo {year} {1981})}\BibitemShut {NoStop}%
\bibitem [{\citenamefont {Ueda}\ \emph {et~al.}(1990)\citenamefont {Ueda},
  \citenamefont {Imoto},\ and\ \citenamefont {Ogawa}}]{ueda_quantum_1990}%
  \BibitemOpen
  \bibfield  {author} {\bibinfo {author} {\bibfnamefont {M.}~\bibnamefont
  {Ueda}}, \bibinfo {author} {\bibfnamefont {N.}~\bibnamefont {Imoto}}, \ and\
  \bibinfo {author} {\bibfnamefont {T.}~\bibnamefont {Ogawa}},\ }\href
  {\doibase 10.1103/PhysRevA.41.3891} {\bibfield  {journal} {\bibinfo
  {journal} {Phys. Rev. A}\ }\textbf {\bibinfo {volume} {41}},\ \bibinfo
  {pages} {3891} (\bibinfo {year} {1990})}\BibitemShut {NoStop}%
\bibitem [{\citenamefont {Drummond}(1987)}]{drummond_unifying_1987}%
  \BibitemOpen
  \bibfield  {author} {\bibinfo {author} {\bibfnamefont {P.~D.}\ \bibnamefont
  {Drummond}},\ }\href {\doibase 10.1103/PhysRevA.35.4253} {\bibfield
  {journal} {\bibinfo  {journal} {Phys. Rev. A}\ }\textbf {\bibinfo {volume}
  {35}},\ \bibinfo {pages} {4253} (\bibinfo {year} {1987})}\BibitemShut
  {NoStop}%
\bibitem [{\citenamefont
  {Fleischhauer}(1998)}]{fleischhauer_quantum-theory_1998}%
  \BibitemOpen
  \bibfield  {author} {\bibinfo {author} {\bibfnamefont {M.}~\bibnamefont
  {Fleischhauer}},\ }\href {\doibase 10.1088/0305-4470/31/2/007} {\bibfield
  {journal} {\bibinfo  {journal} {J. Phys. A-Math. and Gen.}\ }\textbf
  {\bibinfo {volume} {31}},\ \bibinfo {pages} {453} (\bibinfo {year}
  {1998})}\BibitemShut {NoStop}%
\bibitem [{\citenamefont {Sperling}\ \emph {et~al.}(2012)\citenamefont
  {Sperling}, \citenamefont {Vogel},\ and\ \citenamefont
  {Agarwal}}]{sperling_true_2012}%
  \BibitemOpen
  \bibfield  {author} {\bibinfo {author} {\bibfnamefont {J.}~\bibnamefont
  {Sperling}}, \bibinfo {author} {\bibfnamefont {W.}~\bibnamefont {Vogel}}, \
  and\ \bibinfo {author} {\bibfnamefont {G.~S.}\ \bibnamefont {Agarwal}},\
  }\href {\doibase 10.1103/PhysRevA.85.023820} {\bibfield  {journal} {\bibinfo
  {journal} {Phys. Rev. A}\ }\textbf {\bibinfo {volume} {85}},\ \bibinfo
  {pages} {023820} (\bibinfo {year} {2012})}\BibitemShut {NoStop}%
\bibitem [{\citenamefont {van Enk}(2017)}]{vanEnck:2017}%
  \BibitemOpen
  \bibfield  {author} {\bibinfo {author} {\bibfnamefont {S.~J.}\ \bibnamefont
  {van Enk}},\ }\href {http://stacks.iop.org/2399-6528/1/i=4/a=045001}
  {\bibfield  {journal} {\bibinfo  {journal} {J. Phys. Commun.}\ }\textbf
  {\bibinfo {volume} {1}},\ \bibinfo {pages} {045001} (\bibinfo {year}
  {2017})}\BibitemShut {NoStop}%
\bibitem [{\citenamefont {Young}\ \emph {et~al.}(2018)\citenamefont {Young},
  \citenamefont {Sarovar},\ and\ \citenamefont {L\'eonard}}]{Young:2018}%
  \BibitemOpen
  \bibfield  {author} {\bibinfo {author} {\bibfnamefont {S.~M.}\ \bibnamefont
  {Young}}, \bibinfo {author} {\bibfnamefont {M.}~\bibnamefont {Sarovar}}, \
  and\ \bibinfo {author} {\bibfnamefont {F.}~\bibnamefont {L\'eonard}},\ }\href
  {\doibase 10.1103/PhysRevA.97.033836} {\bibfield  {journal} {\bibinfo
  {journal} {Phys. Rev. A}\ }\textbf {\bibinfo {volume} {97}},\ \bibinfo
  {pages} {033836} (\bibinfo {year} {2018})}\BibitemShut {NoStop}%
\bibitem [{\citenamefont {Baragiola}\ \emph {et~al.}(2012)\citenamefont
  {Baragiola}, \citenamefont {Cook}, \citenamefont {Branczyk},\ and\
  \citenamefont {Combes}}]{Baragiola:2012cs}%
  \BibitemOpen
  \bibfield  {author} {\bibinfo {author} {\bibfnamefont {B.~Q.}\ \bibnamefont
  {Baragiola}}, \bibinfo {author} {\bibfnamefont {R.~L.}\ \bibnamefont {Cook}},
  \bibinfo {author} {\bibfnamefont {A.~M.}\ \bibnamefont {Branczyk}}, \ and\
  \bibinfo {author} {\bibfnamefont {J.}~\bibnamefont {Combes}},\ }\href@noop {}
  {\bibfield  {journal} {\bibinfo  {journal} {Phys. Rev. A}\ }\textbf {\bibinfo
  {volume} {86}},\ \bibinfo {pages} {013811} (\bibinfo {year}
  {2012})}\BibitemShut {NoStop}%
\bibitem [{\citenamefont {Jacobs}\ and\ \citenamefont
  {Steck}(2006)}]{Jac.Ste-2006}%
  \BibitemOpen
  \bibfield  {author} {\bibinfo {author} {\bibfnamefont {K.}~\bibnamefont
  {Jacobs}}\ and\ \bibinfo {author} {\bibfnamefont {D.~A.}\ \bibnamefont
  {Steck}},\ }\href@noop {} {\bibfield  {journal} {\bibinfo  {journal}
  {Contemp. Phys.}\ }\textbf {\bibinfo {volume} {47}},\ \bibinfo {pages} {279}
  (\bibinfo {year} {2006})}\BibitemShut {NoStop}%
\bibitem [{\citenamefont {Blais}\ \emph {et~al.}(2004)\citenamefont {Blais},
  \citenamefont {Huang}, \citenamefont {Wallraff}, \citenamefont {Girvin},\
  and\ \citenamefont {Schoelkopf}}]{blais_cavity_2004}%
  \BibitemOpen
  \bibfield  {author} {\bibinfo {author} {\bibfnamefont {A.}~\bibnamefont
  {Blais}}, \bibinfo {author} {\bibfnamefont {R.-S.}\ \bibnamefont {Huang}},
  \bibinfo {author} {\bibfnamefont {A.}~\bibnamefont {Wallraff}}, \bibinfo
  {author} {\bibfnamefont {S.~M.}\ \bibnamefont {Girvin}}, \ and\ \bibinfo
  {author} {\bibfnamefont {R.~J.}\ \bibnamefont {Schoelkopf}},\ }\href@noop {}
  {\bibfield  {journal} {\bibinfo  {journal} {Phys. Rev. A}\ }\textbf {\bibinfo
  {volume} {69}},\ \bibinfo {pages} {062320} (\bibinfo {year}
  {2004})}\BibitemShut {NoStop}%
\bibitem [{\citenamefont {Gross}\ and\ \citenamefont
  {Haroche}(1982)}]{gross_superradiance:_1982}%
  \BibitemOpen
  \bibfield  {author} {\bibinfo {author} {\bibfnamefont {M.}~\bibnamefont
  {Gross}}\ and\ \bibinfo {author} {\bibfnamefont {S.}~\bibnamefont
  {Haroche}},\ }\href@noop {} {\bibfield  {journal} {\bibinfo  {journal} {Phys.
  Rep.}\ }\textbf {\bibinfo {volume} {93}},\ \bibinfo {pages} {301} (\bibinfo
  {year} {1982})}\BibitemShut {NoStop}%
\bibitem [{\citenamefont {Kraus}(1983)}]{kraus_states_1983}%
  \BibitemOpen
  \bibfield  {author} {\bibinfo {author} {\bibfnamefont {K.}~\bibnamefont
  {Kraus}},\ }\href@noop {} {\emph {\bibinfo {title} {States, {Effects}, and
  {Operations}: {Fundamental} {Notions} of {Quantum} {Theory}}}},\ \bibinfo
  {series} {Book}, Vol.\ \bibinfo {volume} {190}\ (\bibinfo  {publisher}
  {Springer},\ \bibinfo {year} {1983})\BibitemShut {NoStop}%
\bibitem [{\citenamefont {Nielsen}\ and\ \citenamefont
  {Chuang}(2010)}]{nielsen_quantum_2010}%
  \BibitemOpen
  \bibfield  {author} {\bibinfo {author} {\bibfnamefont {M.~A.}\ \bibnamefont
  {Nielsen}}\ and\ \bibinfo {author} {\bibfnamefont {I.~L.}\ \bibnamefont
  {Chuang}},\ }\href@noop {} {\emph {\bibinfo {title} {Quantum computation and
  quantum information}}},\ Book\ (\bibinfo  {publisher} {Springer},\ \bibinfo
  {year} {2010})\BibitemShut {NoStop}%
\end{thebibliography}%

\end{document}